\documentclass[aps,prl,twocolumn,10pt,nobalancelastpage,superscriptaddress,longbibliography]{revtex4-1}
\usepackage[pdftex]{graphicx}
\usepackage{amsmath}
\usepackage{amssymb}
\usepackage{amsthm}
\usepackage[dvipsnames]{color}
\usepackage[colorlinks=True,linkcolor=red,citecolor=blue,urlcolor=blue]{hyperref}
\usepackage{braket}
\usepackage{bm}
\usepackage{pdfpages}
\makeatletter
\AtBeginDocument{\let\LS@rot\@undefined}
\makeatother

\newcommand{\pd}{\phantom\dagger}

\newcommand{\typ}{\mathrm{typ}}
\newcommand{\yt}{y_{0,\mathrm{typ}}^{{\phantom\dagger}}}
\newcommand{\nei}{\mathsf{N}}
\newcommand{\ve}{\Omega}
\newcommand{\comment}[1]{}

\newcommand{\mytitle}{Localisation on certain graphs with strongly correlated disorder}
\begin{document}

\title{\mytitle}
\author{Sthitadhi Roy}
\email{sthitadhi.roy@chem.ox.ac.uk}
\affiliation{Physical and Theoretical Chemistry, Oxford University, South Parks Road, Oxford, OX1 3QZ, United Kingdom}
\affiliation{ Rudolf Peierls Centre for Theoretical Physics, Clarendon Laboratory,
Oxford University, Parks Road, Oxford OX1 3PU, United Kingdom}

\author{David E. Logan}
\email{david.logan@chem.ox.ac.uk}
\affiliation{Physical and Theoretical Chemistry, Oxford University, South Parks Road, Oxford, OX1 3QZ, United Kingdom}
\affiliation{Department of Physics, Indian Institute of Science,
Bangalore 560012, India}

\begin{abstract}
Many-body localisation in interacting quantum systems can be cast as a disordered hopping problem on the underlying Fock-space graph. A crucial feature of the effective Fock-space disorder is that the Fock-space site energies are strongly correlated -- maximally so for sites separated by a finite distance on the graph. Motivated by this, and to understand the effect of such correlations more fundamentally, we study Anderson localisation on Cayley trees and random regular graphs, with maximally correlated disorder. Since  such correlations suppress short distance fluctuations in the disorder potential, one might naively suppose they disfavour localisation.  We find however  that there exists an Anderson transition, and indeed that localisation is more robust in the sense that the critical disorder scales with graph connectivity $K$ as $\sqrt{K}$, in marked contrast to $K\ln K$ in the uncorrelated case. This scaling is argued to be intimately connected to the stability of many-body localisation. Our analysis centres on an exact recursive formulation for the local propagators as  well as a self-consistent mean-field theory; with results corroborated using exact  diagonalisation.
\end{abstract}

\maketitle

Disorder-induced localisation of non-interacting quantum particles -- the phenomenon of Anderson localisation (AL) -- has been one of the most profound discoveries in physics~\cite{anderson1958absence}. Its robustness to interactions in quantum many-body systems has lately been a major research theme, under the banner of many-body localisation (MBL)~\cite{gornyi2005interacting,basko2006metal,oganesyan2007localisation,pal2010many} (see Refs.~\cite{nandkishore2015many,alet2018many,abanin2019colloquium} for reviews and further references). MBL systems fall outside the paradigm of conventional statistical mechanics allowing for novel quantum phases, and are thus of fundamental interest.

Efforts to understand the MBL phase and the accompanying MBL transition have ranged from extensive numerical studies~\cite{kjall2014many,luitz2015many,alet2018many} and phenomenological treatments~\cite{vosk2015theory,potter2015universal,goremykina2019analytically,dumitrescu2018kosterlitz,morningstar2019renormalization,morningstar2020manybody} to studying the problem directly on the Fock space~\cite{logan1990quantum,altshuler1997quasiparticle,MonthusGarel2010PRB,pietracaprina2016forward,logan2019many,roy2018exact,roy2018percolation,roy2019self,pietracaprina2019hilbert,ghosh2019manybody,roy2020fock}.  One virtue of the latter is that the problem  can be cast as a disordered hopping problem on the Fock-space graph, thus offering the prospect of exploiting techniques and understandings developed for AL. However, MBL on Fock space is fundamentally different from conventional 
AL on high-dimensional graphs, due to the presence of  \emph{maximal correlations} in the effective Fock-space disorder:  
the statistical correlation between two Fock-space site energies, scaled by their variance, approaches its maximum value of
unity in the thermodynamic limit, for any pair separated by a finite Hamming distance on the Fock-space graph. 
This was found  to be a necessary condition for MBL to exist~\cite{roy2020fock}.

Motivated by this, here we ask a fundamental question: what is the fate of AL on random graphs with maximally correlated disorder?  In parallel to the case of Fock-space disorder, the correlation between the disordered site energies of 
any two sites separated by a finite distance on the graph takes it maximum value in the thermodynamic limit. 
In suppressing fluctuations in the site-energies, one might naively suppose these correlations would strongly favour delocalisation; indeed it is not \emph{a priori} obvious that a localised phase must exist in such a case. Nevertheless, 
not only do we find inexorably a localised phase and an Anderson transition, but also that the scaling of the critical disorder with graph connectivity is qualitatively different to that for the standard model with uncorrelated disorder. These models thus introduce a novel class of AL problems with intimate connections to the problem of MBL  on Fock space.

Concretely, we consider a disordered tight-binding model on a rooted Cayley tree (as well as on random regular graphs (RRG) which are locally tree-like). For uncorrelated disorder, such models have served as archetypes for studying a range of phenomena such as localisation transitions, multifractality, and glassy dynamics on complex high-dimensional graphs~\cite{abou-chacra1973self,chalker1990anderson,luca2014anderson,altshuler2016multifractal,tikhonov2016anderson,garciamata2017scaling,sonner2017multifractality,biroli2018delocalization,kravtsov2018nonergodic,tikhonov2019critical,savitz2019anderson,garciamata2020two,tarzia2020manybody,biroli2017delocalized,biroli2020anomalous,detomasi2020subdiffusion}.
The model Hamiltonian is
\begin{equation}
    H = \Gamma\sum_{\braket{i,j}}[\ket{i}\bra{j}+\mathrm{h.c.}]+W\sum_{i}\epsilon_i\ket{i}\bra{i}
    \label{eq:ham}
\end{equation}
in the position basis $\{\ket{i}\}$, where $\braket{i,j}$ denotes a sum over nearest neighbour pairs. We denote the branching number of the tree by $K$ and the total number of generations in a finite-sized tree by $L$; the total number of sites in the tree is $N\sim K^L$. The set of correlated random site-energies, $\{\epsilon_i\}$, is fully specified by a $N$-dimensional joint distribution. To mimic the case of many-body systems on Fock space~\cite{welsh2018simple,logan2019many,roy2020fock}, we take these distributions to be multivariate Gaussians, $\mathcal{N}(\bm{0},\mathbf{C})$, characterised completely by the covariance matrix $\mathbf{C}$~\cite{roy2020fock}. Taking a cue from disordered  interacting local Hamiltonians, we consider the matrix elements $C_{ij}$ to depend only on the  distance $\ell_{ij}$ between a pair of sites. To impose the maximally correlated limit, we consider
\begin{equation}
    C_{ij}^{\pd} = \braket{\epsilon_i^{\pd}\epsilon_j^{\pd}}=f(\ell_{ij}/L);~~~\lim_{x\to0}f(x)=1.
    \label{eq:maxcorr}
\end{equation}
The functional form of $f$ does not qualitatively affect our results, but for concreteness in numerical calculations we take $C_{ij} = \exp[-\ell_{ij}/\lambda L]$ with $\lambda=1$~\footnote{The algorithm for constructing the correlated energies is described in the supplementary material~\cite{supp}}. The choice of the argument of $f$ is motivated by the form of correlations in the Fock-space disorder of disordered many-body systems; for $p$-local Hamiltonians the analogous $f$ was shown to be a $p^\mathrm{th}$-order polynomial of $\ell_{ij}/\ln N_\mathcal{H}$, $N_\mathcal{H}$ being the Fock-space dimension~\cite{roy2020fock}.

\begin{figure*}
    \centering
    \includegraphics[width=\linewidth]{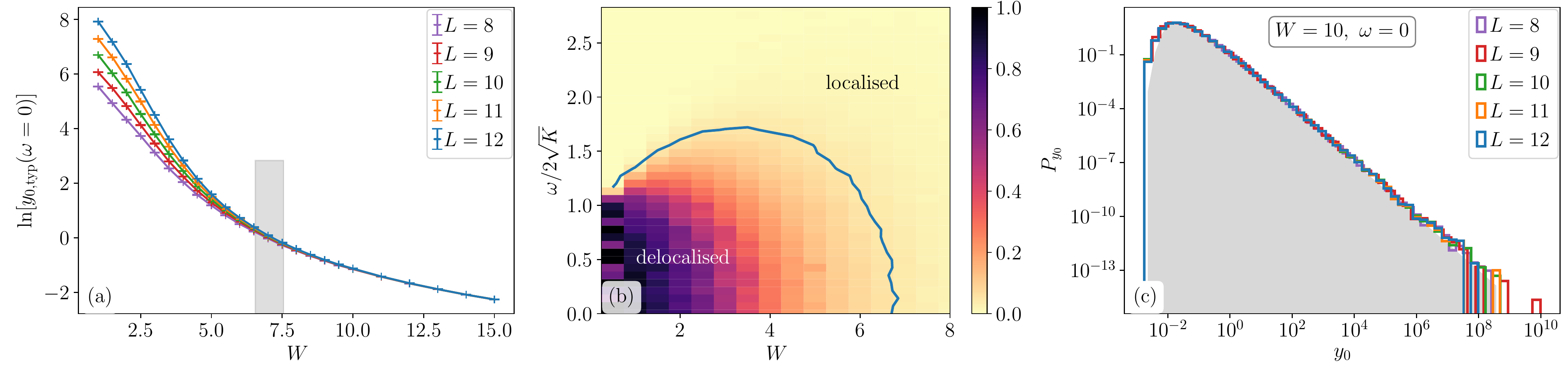}
    \caption{
            For a $K=2$ rooted Cayley tree, numerical results from the exact recursion method. 
            (a) The typical $y_{0,\mathrm{typ}}$ at $\omega=0$, \emph{vs} disorder strength $W$ (with $\Gamma \equiv 1$)
            for different total generation numbers $L$. For $W>W_c$, $y_{0,\mathrm{typ}}$ is independent of $L$; while for $W<W_c$ it grows  with $L$, indicating a divergence in the thermodynamic limit. The critical $W_c$ subject to errorbars is  the grey shaded region, estimated by positing $y_{0,\mathrm{typ}}=A+B N^\beta$; $\beta=0$ implies a localised phase and its deviation from 0 the onset of 
delocalisation. 
            (b) Colour-map of $\beta$ as a function of $(W,\omega)$. The blue line shows the contour $\beta=0.05$ as an estimate of the critical line (mobility edges); the value is chosen in accordance with the errorbars in $\beta$. For 
$\omega=0$ our best estimate is $W_c(\omega=0)\simeq 6.8$. 
            (c) Distribution of $y_0^{[L]}$ in the localised phase. Data are well converged for different $L$, and in excellent agreement with a L\'evy distribution shown by the grey shaded region. Statistics are obtained over 
$5$$\times$$10^{4}$ disorder realisations.}
    \label{fig:recursive-tree}
\end{figure*}

Our analysis centres on the local Feenberg self-energy $S_i(\omega)\equiv X_i(\omega)-i\Delta_i(\omega)$, defined via the local propagator as $G_i(\omega)=[\omega^+ - \epsilon_i - S_i(\omega)]^{-1}$ with $\omega^+=\omega+i\eta$ ($\eta =0^{+}$). 
We focus on the imaginary part of the self-energy, $\Delta_i(\omega)$, as it serves as a probabilistic order parameter for a localisation transition.  Physically, $\Delta_i(\omega)$ gives the rate of loss of probability from site $i$ into states of energy $\omega$. In a delocalised phase $\Delta_i(\omega)$ is finite,  whereas in a localised phase  it vanishes $\propto\eta$ (with $y_{i}(\omega)=\Delta_{i}(\omega)/\eta$ finite), both with unit probability. These 
characteristics of $\Delta_i(\omega)$ have long been used successfully to understand Anderson transitions~\cite{anderson1958absence,economous1972existence,abou-chacra1973self,ThoulessReview1974,Licciardello+Economou1975,logan1985anderson,*DELPGWPRB1987}; 
and, more recently, MBL transitions on Fock space~\cite{logan2019many,roy2019self,roy2020fock}.

We focus on the self-energy of the root site ($i=0$) of the rooted Cayley tree. $S_{0}(\omega)$ is given exactly by
\begin{equation}
    S_0^{\pd}(\omega) = \Gamma^2\sum_{i_1\in\nei[0]}[\omega^+-W\epsilon_{i_1}^{\pd}-S_{i_1}^{(0)}]^{-1},
    \label{eq:S0def}
\end{equation}
with the sum over all sites in the first generation, and $S_{i_1}^{(0)}$ the self-energy of site $i_1$ with the root site removed. One could in principle now approximate the self-energy on the right-hand side of  Eq.~\eqref{eq:S0def} by a typical $S_\typ$, and obtain the distribution of $S_0$ self-consistently~\cite{logan2019many,roy2019self,roy2020fock}.
Here however we go far beyond such a treatment, addressing Eq.~\eqref{eq:S0def} to arbitrarily high orders  via an exact recursive method. We first sketch the formulation, focussing on the localised phase, in particular its stability and self-consistency; whence the quantity of interest is $y_0(\omega)$.

From Eq.~\eqref{eq:S0def}, $y_0^{\pd}(\omega)$ can be expressed as 
\begin{equation}
    y_0^{\pd} = \sum_{i_1\in\nei[0]}\frac{\Gamma^2}{\ve_{i_1}^2}\left[1+y_{i_1}^{(0)}\right];~\ve_{i_1}^{\pd} = \omega-W\epsilon_{i_1}^{\pd}-X_{i_1}^{(0)}.
    \label{eq:y0}
\end{equation}
This is a recursion relation, which can be iterated as
\begin{equation}
    y_0^{\pd} = \sum_{i_1\in\nei[0]}\frac{\Gamma^2}{\ve_{i_1}^2}\left[1+\sum_{i_2\in\nei[i_1]}\frac{\Gamma^2}{\ve_{i_2}^2}\left[1+\sum_{i_3\in\nei[i_2]}\frac{\Gamma^2}{\ve_{i_3}^2}[1+\cdots\right.\right..
    \label{eq:y0-recursion}
\end{equation}
In Eq.~\eqref{eq:y0-recursion}, for any site $i_n$ on generation $n$ of the tree,  $\ve_{i_n}=\omega-W\epsilon_{i_n}-X_{i_n}^{(i_{n-1})}(\omega)$, with $X_{i_n}^{(i_{n-1})}$ the real part of the self-energy of site $i_n$ with its (unique) neighbour $i_{n-1}$ on the previous generation removed. As for the imaginary part of the self-energy, a recursion relation for the real part can also be derived from Eq.~\eqref{eq:S0def}. This leads to a recursion relation for 
$\ve_{i_n}$,
\begin{equation}
    \ve_{i_n}^{\pd} = \omega-W\epsilon_{i_n}^{\pd}-\sum_{i_{n+1}\in\nei[i_{n}]}\frac{\Gamma^2~~~}{\ve_{i_{n+1}}^2},
    \label{eq:ve-recursion}
\end{equation}
with the boundary condition $\ve_{i_L}=\omega-W\epsilon_{i_L}$ for a  tree with $L$ generations. Eqs.~\eqref{eq:y0-recursion},\eqref{eq:ve-recursion} comprise the complete set of recursion relations required to compute $y_0(\omega)$ to all orders. We now make key conceptual points about the stability of the localised phase or lack thereof, and describe our results.

Note that by evaluating $y_0(\omega)$ using Eq.~\eqref{eq:y0-recursion} for many disorder realisations, one can generate its entire distribution $P_{y_0}$, and also compute its typical value via $\ln \yt = \int dy_0~P_{y_0}(y_0)\ln y_0$. 
A stable localised phase is indicated by $\yt$ taking a finite value independent of system size; whereas the delocalised phase is identified via a systematic growth of $\yt$ with system size, such that it diverges in the thermodynamic limit. The disorder strength separating these two behaviours, if present, is the critical disorder. Numerical results for the localisation phase digaram so obtained for a $K=2$ Cayley tree with maximally correlated disorder are shown in Fig.~\ref{fig:recursive-tree}. Considering the band centre $\omega=0$ as an example (panel (a)), $\ln \yt$ is independent of $L$ for $W>W_c$ whereas it diverges with $L$ for $W<W_c$; thus showing that a localisation transition is indeed present in the model. The phase diagram similarly obtained in the entire $\omega$-$W$ plane is given in  Fig.~\ref{fig:recursive-tree}(b), which shows the presence of mobility edges in the spectrum. Finally,  Fig.~\ref{fig:recursive-tree}(c), the distribution of $y_0$ is shown for a representative disorder in the localised phase, and shows excellent agreement with a L\'evy distribution characteristic of a localised phase, $P_{y_0}(y_0) = \sqrt{\kappa/\pi}~y_0^{-3/2}e^{-\kappa/y_0}$ with scale parameter $\kappa$.

\begin{figure}
    \centering
    \includegraphics[width=\linewidth]{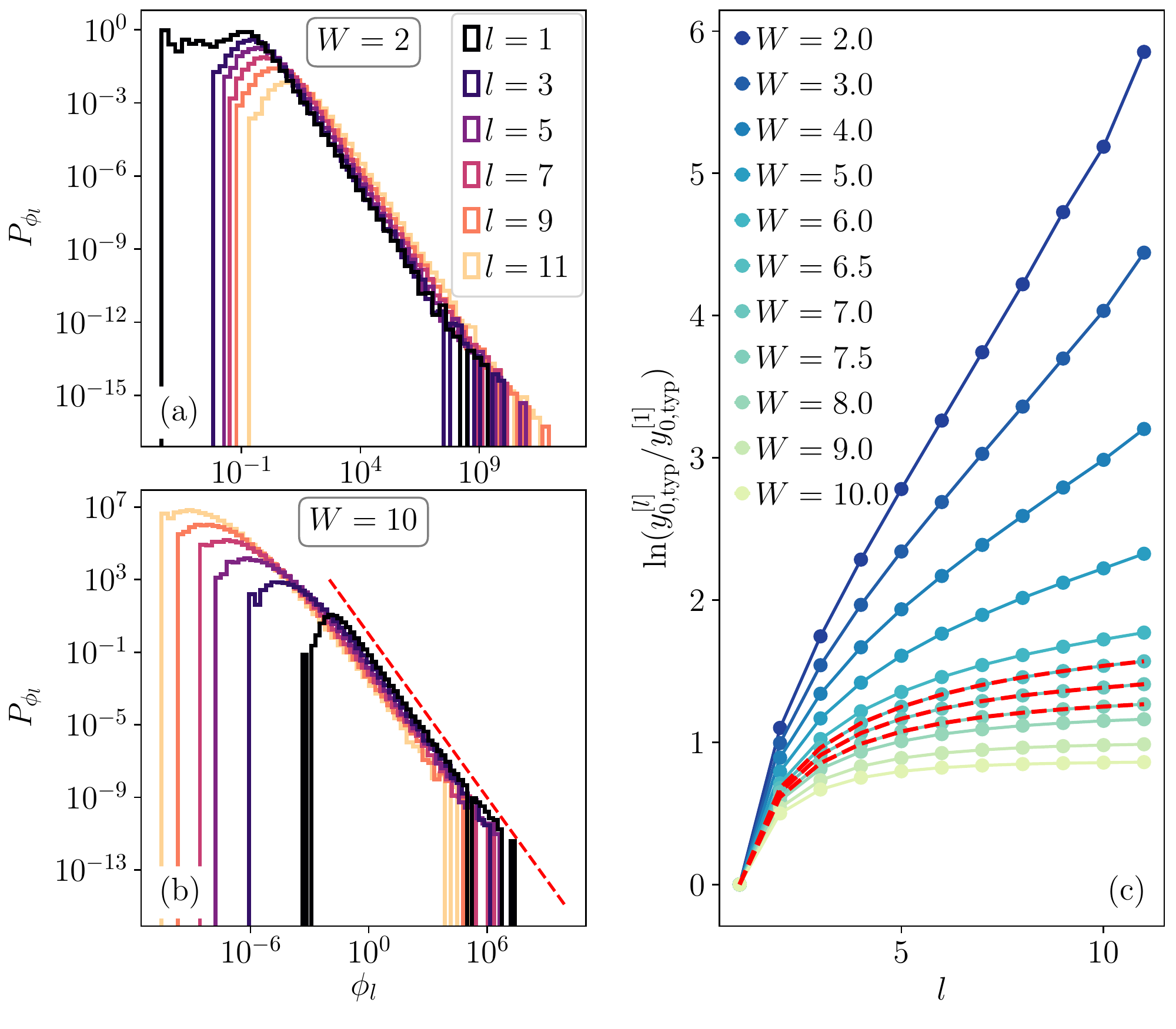}
    \caption{Convergence of the series for $y_0$ (Eq.~\eqref{eq:y0-series}) for the $K=2$ Cayley tree. (a)-(b) Distributions $P_{\phi_l}$ in delocalised and localised phases, for different $l$; evolution with $l$ is qualitatively different in the two phases. Dashed line for localised phase shows L\'evy tail, slope $-\tfrac{3}{2}$. (c) The typical value of the series summed to $l$ terms (normalised by the $l=1$ value)  \emph{vs} $l$. Results for $L=12$, $\omega=0$, and $5$$\times$$10^{4}$ realisations. Red lines show $W$ values lying in the critical regime shown in Fig.~\ref{fig:recursive-tree}(a). }
    \label{fig:recursive-conv}
\end{figure}

The stability of the localised phase can also be understood as the convergence of the recursion relation in Eq.~\eqref{eq:y0-recursion}. The series for $y_0$ can be organised as 
\begin{equation}
    y_0^{\pd} = \sum_{l=1}^\infty \phi_l^{\pd}; ~\phi_l^{\pd}=
    \sum_{i_1\in\nei[0]}\frac{\Gamma^2}{\ve_{i_1}^2}
    \sum_{i_2\in\nei[i_1]}
    \frac{\Gamma^2}{\ve_{i_2}^2}
    \cdots\sum_{i_l\in\nei[i_{l-1}]}\frac{\Gamma^2}{\ve_{i_l}^2},
    \label{eq:y0-series}
\end{equation}
with $\phi_l$ the total contribution to $y_0$ from all sites on the $l^\mathrm{th}$ generation. Diagrammatically, it is the total contribution to $y_0$ from all $K^l$ paths of length $2l$, each of which goes from the root site to a unique site in the $l^\mathrm{th}$ generation and retraces itself back to the root site~\footnote{On a tree, there exists a unique shortest path between any pair of sites. For a site on generation $l$, the length of the corresponding path between it and the root site is $l$.}. For the series in Eq.~\eqref{eq:y0} to converge in the thermodynamic limit, $\phi_l$ must decrease sufficiently fast with increasing $l$. This suggests that the distributions $P_{\phi_l}$ of $\phi_l$, should evolve with $l$ in a qualitatively different manner in the delocalised and localised phases.  Calculating $P_{\phi_l}$ shows that this is indeed so, as shown in Fig.~\ref{fig:recursive-conv}(a)-(b). For strong disorder (localised phase), the vast bulk of the distribution shifts rapidly to smaller values with increasing $l$, while in the delocalised phase the support of the $P_{\phi_l}$ moves to larger values with increasing $l$. This is itself  indicative of the convergence of the series in the localised phase and otherwise in the delocalised. To further quantify the convergence, one can define 
$y_0^{\scriptstyle{[l]}}\equiv\sum_{n=1}^l\phi_n$ and study its typical value, $y_{0,\mathrm{typ}}^{\scriptstyle{[l]}}$, as a function of $l$ and $W$. Representative results at $\omega=0$ are shown in Fig.~\ref{fig:recursive-conv}(c). For weak disorder, $y_{0,\mathrm{typ}}^{\scriptstyle{[l]}}$ grows rapidly with $l$, whereas for strong disorder it saturates to its converged value in the localised phase; again clearly showing the presence of a localisation transition.

Two further remarks should be made. First,  the recursive formulation also treats the \emph{real} parts of all self-energies exactly. One can however make the simplifying approximation of neglecting them -- Anderson's `upper limit approximation'~\cite{anderson1958absence,abou-chacra1973self}. For the tree with correlated disorder this approximation again predicts the presence of a transition, albeit naturally at a higher $W_c$~\cite{supp}. Second, the terms appearing in the series in Eq.~\eqref{eq:y0-series} but with $X_{i_{n}}^{\scriptstyle{(i_{n-1})}}=0$ (i.e.\ $\ve_{i_{n}}\equiv\omega-W\epsilon_{i_{n}}$) are precisely those appearing in the Forward Approximation~\cite{pietracaprina2016forward}. By including the contribution of non-local propagators to the local propagator in an exact, \emph{fully} renormalised 
fashion, the recursive formulation is a significant technical advance.

Correlations in the $\epsilon_i$'s preclude an exact analytic solution for the distribution of $y_0$ from Eq.~\eqref{eq:y0-recursion}. One can nevertheless perform a self-consistent mean-field calculation analytically at leading order in the renormalised perturbation series~\cite{logan2019many,roy2020fock,roy2019self} (here illustrated for $\omega=0$). Here $y_0$ depends only on the site energies of its neighbours, $\{i_1\}$. Since $\ell_{0i_1}=1$, the maximally correlated limit implies the conditional distribution $P(\epsilon_{i_1}|\epsilon_0)=\delta(\epsilon_{i_1}-\epsilon_0)$ in the thermodynamic limit. The distribution of $y_0$ can thus be simply calculated as $P_{y_{0}}(y_{0})=\int d\epsilon_{0}~P(\epsilon_0^{\pd})~\delta\big(y_0^{\pd}-K\Gamma^2(1+\yt)/[W^2\epsilon_0^2]\big)$. Since the univariate distribution $P(\epsilon_0)$ is a standard Normal, this yields $P_{y_0}(y_0,\yt) = \sqrt{\kappa/\pi}~e^{-\kappa/y_0}y_0^{-3/2}$ where $\kappa = K(1+\yt)\Gamma^2/2W^2$. Remarkably and reassuringly, the distribution indeed has the  L\'evy form, just as 
obtained numerically by summing the entire series (Fig.~\ref{fig:recursive-tree}(c)).

Self-consistency can now be imposed by requiring $\ln\yt=\int dy_0P_{y_0}(y_0,\yt)\ln y_0$; the solution of which is
$\yt=2e^\gamma K\Gamma^2(W^2-2e^\gamma K\Gamma^2)^{-1}$, with $\gamma$ the Euler-Mascheroni constant. Since $y_0$ is necessarily non-negative, self-consistency of the localised phase requires $W\ge W_c$, with~\footnote{While this analysis focuses on the localised phase, self-consistency for the delocalised phase commensurately breaks down at the same $W_{c}$ as in Eq.~\eqref{eq:Wc-sc}~\cite{logan2019many,roy2019self}.}
\begin{equation}
    W_c^{\pd} = \sqrt{2}e^{\gamma/2}\Gamma\sqrt{K}.
    \label{eq:Wc-sc}
\end{equation}
This $W_{c}\propto \sqrt{K}$ scaling is qualitatively different from that arising for uncorrelated disorder, where $W_c\propto K\ln K$~\cite{abou-chacra1973self}; and stems intrinsically from the maximal correlations in the disorder.

We turn now to results arising for RRGs, via exact diagonalisation (ED) of tight-binding Hamiltonians Eq.~\eqref{eq:ham} with maximally correlated disorder Eq.~\eqref{eq:maxcorr}. Our motivation here is twofold. First, while results above were for a rooted Cayley tree, we expect them to hold qualitatively for other random graphs. Second, it is important to corroborate the results with other independent measures of localisation. Cayley trees are not moreover readily amendable to ED, since a finite fraction of sites live on the boundary; this issue is sidestepped by considering RRGs, which are locally tree-like but contain long loops.

In the following we consider RRGs with a coordination number $Z=K+1 =3$; denoting the total number of sites in the RRG by $N$. In accordance with the form of the covariance matrix for the Cayley tree, we take $C_{ij}=\exp[-\ell_{ij}\ln K/\ln N]$. The quantities studied will be the level spacing ratios, and $\Delta_i$ computed directly. We focus on the middle of the spectrum ($\omega =0$) and consider 50-100 eigenstates therein.

\begin{figure}
    \centering
    \includegraphics[width=\linewidth]{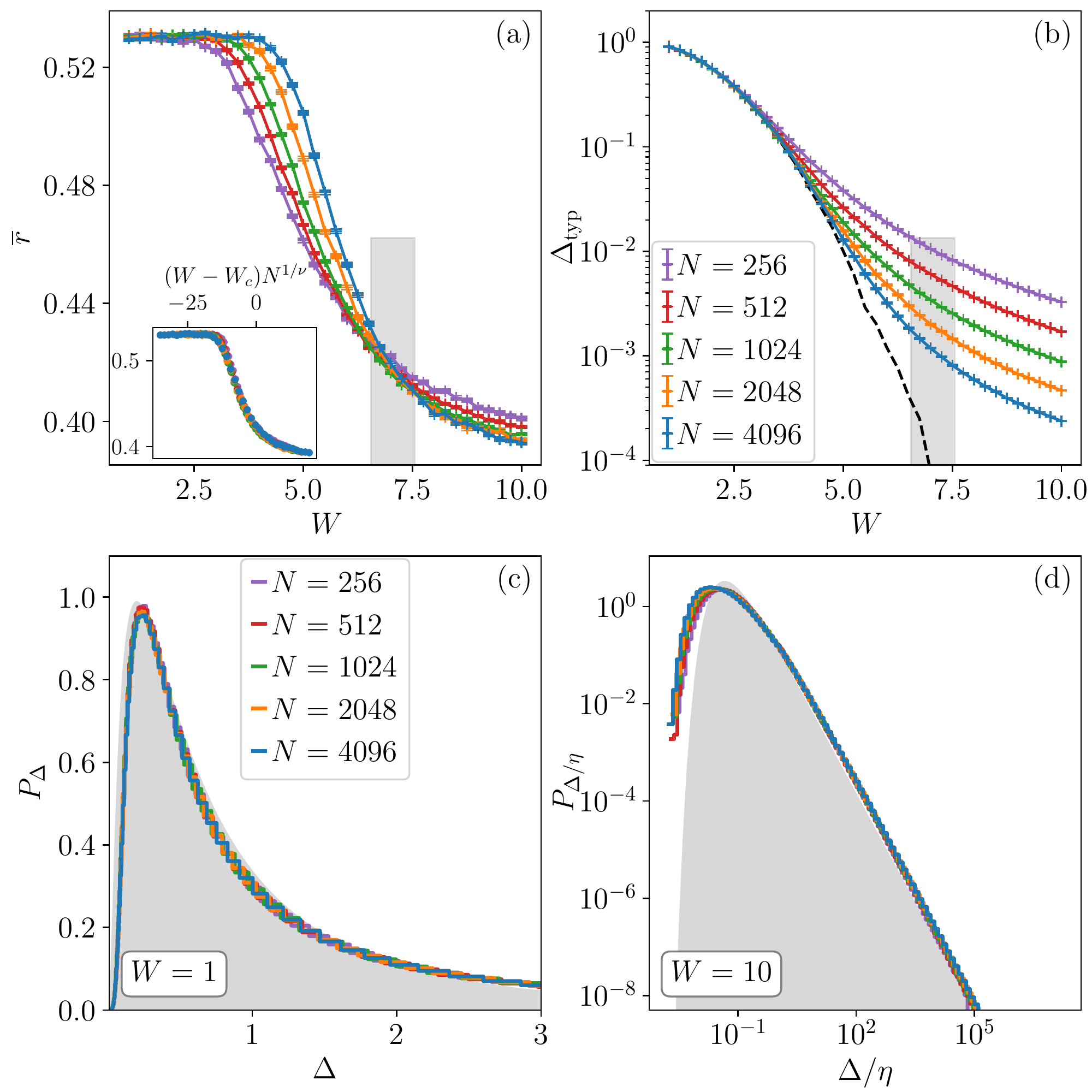}
    \caption{
        ED results for a $K=2$ RRG with maximally correlated disorder. (a) Mean level spacing ratio \emph{vs} $W$ shows a crossing for different $N$. Data collapse onto a common function of $(W-W_c)N^{1/\nu}$ yields $W_c\simeq 6.8$ and $\nu\simeq4.6$ (inset). (b) Typical value $\Delta_{\typ}(\omega =0)$ computed exactly from Eq.~\eqref{eq:Delta-RRG}. In the delocalised [localised] phase it is independent of [decays with] $N$. Dashed line shows extrapolation to $N\to\infty$. Grey shaded regions in (a), (b) denote the estimated critical region. (c)-(d) Distributions of $\Delta$ and $y=\Delta/\eta$ in the delocalised and localised phases respectively. Grey shaded regions show best fits to Log-Normal and L\'evy distributions respectively.}
    \label{fig:rrg-ed}
\end{figure}

For an ordered set of eigenvalues $\{E_n\}$, the level spacing ratio is $r_n=\min[s_n,s_{n+1}]/\max[s_n,s_{n+1}]$ with $s_n=E_n-E_{n-1}$. In an ergodic phase the distribution of $r_n$ follows the Wigner-Dyson surmise with mean  $\overline{r}\simeq0.53$, while in a localised phase the distribution is Poisson with $\overline{r}\simeq 0.386$.  Results for $\overline{r}$ \emph{vs} $W$ are shown in Fig.~\ref{fig:rrg-ed}(a), and show clearly a localisation transition. A scaling collapse of the data for various $N$ onto a common function of $(W-W_c)N^{1/\nu}$ yields a critical disorder strength of $W_c\simeq 6.8$ and $\nu\simeq 4.6$. Note that the $W_c$ estimated is remarkably close to that obtained above numerically
for the $K=2$ Cayley tree.

From the set of exact eigenvalues $\{E_n\}$ and eigenstates $\{\ket{\psi_n}\}$, $\Delta_i(\omega)$ can be computed as
\begin{equation}
    \Delta_i(\omega) = \mathrm{Im}[G_i^{-1}(\omega)]-\eta,~~G_i=\sum_n\frac{\vert\braket{\psi_n|i}\vert^2}{\omega+i\eta-E_n}.
    \label{eq:Delta-RRG}
\end{equation}
As $\Delta$ is finite with unit probability in the delocalised phase, $\Delta_\mathrm{typ}$ should converge to a finite value with increasing $N$; while in a localised phase  $\Delta \propto\eta$ vanishes with unit probability, so $\Delta_\mathrm{typ}$ should decrease with $N$. This behaviour is indeed found, see Fig.~\ref{fig:rrg-ed}(b). To estimate numerically the critical $W_{c}$, we posit $\Delta_\mathrm{typ}=\Delta_{\mathrm{typ},N\to\infty}+ a/N^\beta$ and extrapolate the data to the thermodynamic limit. As shown in Fig.~\ref{fig:rrg-ed}(b), the vanishing of $\Delta_{\mathrm{typ},N\to\infty}$ gives a $W_c$ consistent with that obtained from level statistics. In the localised phase, the distribution of $y=\Delta/\eta$ is again in very good agreement with a L\'evy distribution (see Fig.~\ref{fig:rrg-ed}(d)).
In the delocalised phase by contrast, $\Delta$ is qualitatively different, and appears to be log-normally distributed (Fig.~\ref{fig:rrg-ed}(c)).

As above, whether for a Cayley tree or RRG, we find a one-parameter L\'evy distribution for $y=\Delta/\eta$ in the localised phase. Importantly, it is thus universal: distributions for different  $W>W_{c}$ can be collapsed onto a universal form by scaling the self-energy as $y/y_\mathrm{typ}$~\cite{supp}. Further, the distribution can be directly connected to that of wavefunction amplitudes, the moments of which (via generalised IPRs) probe the divergence of the localisation length, $\xi$, as $W\to W_{c}$~\cite{supp}. Within our mean-field theory, we find $\xi\sim(W-W_c)^{-1}$ with an exponent of $1$.

We turn now to the $K\to\infty$ limit. For any one-body problem to remain well-defined in this limit, the hopping must be rescaled as $\Gamma = \Gamma_\ast/\sqrt{K}$. The mean-field theory then yields a finite critical  $W_c=\sqrt{2}e^{\gamma/2}\Gamma_\ast$; in stark contrast to the case of uncorrelated disorder where, despite rescaling  $\Gamma$, $W_{c}/\Gamma_\ast\propto\sqrt{K}\ln K$ thus precludes localisation as $K\to\infty$. For MBL on Fock space, in a system containing $L$ real-space sites, the effective connectivity on the Fock-space graph scales as $K\sim L$, and the effective Fock-space disorder as $W_\mathrm{FS}\sim\sqrt{L}W_{\mathrm{t}}$ (with $W_{\mathrm{t}}\sim\mathcal{O}(1)$)~\cite{roy2020fock,logan2019many}. Rescaling all energies by $\sqrt{L}$, as required to attain a well-defined  thermodynamic limit $L\to \infty$,
again leads~\cite{roy2020fock} to a finite critical $W_{\mathrm{t},c}$, in direct parallel to the $K\to\infty$ limit of the present problem. The existence of an MBL phase thus provides an indirect but complementary
argument for the $\sqrt{K}$ scaling of $W_c$.

In summary, we have studied AL on Cayley trees and RRGs with maximally correlated on-site disorder, mimicking the effective Fock-space disorder of MBL systems. While such correlations might be thought to disfavour localisation by suppressing site-energy fluctuations, we find both that an Anderson transition is present, and that scaling of the critical disorder with graph connectivity is qualitatively different from that of uncorrelated disorder, with correlations favouring localisation. Our results address a new class of AL problems, and shed light on the crucial role played by correlations in Fock-space disorder in stabilising MBL. Many questions arise as to what further aspects of MBL can be captured by AL problems with maximally correlated disorder. One such is the multifractal character of wavefunctions, and its possible connection to the anomalous statistics of MBL wavefunctions on Fock space;
and our preliminary results indeed suggest the presence of multifractal eigenstates on RRGs. Looking further afield, understanding the effect of maximal correlations on glassy dynamics on such graphs is also immanently important.

\begin{acknowledgments}
We thank J.~T.~Chalker, A.~Duthie, and A.~Lazarides for useful discussions and comments on the manuscript. This work was in part supported by EPSRC Grant No. EP/S020527/1.
\end{acknowledgments}

\bibliography{refs}

\begin{thebibliography}{53}%
\makeatletter
\providecommand \@ifxundefined [1]{%
 \@ifx{#1\undefined}
}%
\providecommand \@ifnum [1]{%
 \ifnum #1\expandafter \@firstoftwo
 \else \expandafter \@secondoftwo
 \fi
}%
\providecommand \@ifx [1]{%
 \ifx #1\expandafter \@firstoftwo
 \else \expandafter \@secondoftwo
 \fi
}%
\providecommand \natexlab [1]{#1}%
\providecommand \enquote  [1]{``#1''}%
\providecommand \bibnamefont  [1]{#1}%
\providecommand \bibfnamefont [1]{#1}%
\providecommand \citenamefont [1]{#1}%
\providecommand \href@noop [0]{\@secondoftwo}%
\providecommand \href [0]{\begingroup \@sanitize@url \@href}%
\providecommand \@href[1]{\@@startlink{#1}\@@href}%
\providecommand \@@href[1]{\endgroup#1\@@endlink}%
\providecommand \@sanitize@url [0]{\catcode `\\12\catcode `\$12\catcode
  `\&12\catcode `\#12\catcode `\^12\catcode `\_12\catcode `\%12\relax}%
\providecommand \@@startlink[1]{}%
\providecommand \@@endlink[0]{}%
\providecommand \url  [0]{\begingroup\@sanitize@url \@url }%
\providecommand \@url [1]{\endgroup\@href {#1}{\urlprefix }}%
\providecommand \urlprefix  [0]{URL }%
\providecommand \Eprint [0]{\href }%
\providecommand \doibase [0]{http://dx.doi.org/}%
\providecommand \selectlanguage [0]{\@gobble}%
\providecommand \bibinfo  [0]{\@secondoftwo}%
\providecommand \bibfield  [0]{\@secondoftwo}%
\providecommand \translation [1]{[#1]}%
\providecommand \BibitemOpen [0]{}%
\providecommand \bibitemStop [0]{}%
\providecommand \bibitemNoStop [0]{.\EOS\space}%
\providecommand \EOS [0]{\spacefactor3000\relax}%
\providecommand \BibitemShut  [1]{\csname bibitem#1\endcsname}%
\let\auto@bib@innerbib\@empty
\bibitem [{\citenamefont {Anderson}(1958)}]{anderson1958absence}%
  \BibitemOpen
  \bibfield  {author} {\bibinfo {author} {\bibfnamefont {P.~W.}\ \bibnamefont
  {Anderson}},\ }\bibfield  {title} {\enquote {\bibinfo {title} {Absence of
  diffusion in certain random lattices},}\ }\href {\doibase
  10.1103/PhysRev.109.1492} {\bibfield  {journal} {\bibinfo  {journal} {Phys.
  Rev.}\ }\textbf {\bibinfo {volume} {109}},\ \bibinfo {pages} {1492--1505}
  (\bibinfo {year} {1958})}\BibitemShut {NoStop}%
\bibitem [{\citenamefont {Gornyi}\ \emph {et~al.}(2005)\citenamefont {Gornyi},
  \citenamefont {Mirlin},\ and\ \citenamefont
  {Polyakov}}]{gornyi2005interacting}%
  \BibitemOpen
  \bibfield  {author} {\bibinfo {author} {\bibfnamefont {I.~V.}\ \bibnamefont
  {Gornyi}}, \bibinfo {author} {\bibfnamefont {A.~D.}\ \bibnamefont {Mirlin}},
  \ and\ \bibinfo {author} {\bibfnamefont {D.~G.}\ \bibnamefont {Polyakov}},\
  }\bibfield  {title} {\enquote {\bibinfo {title} {Interacting electrons in
  disordered wires: Anderson localization and low-${T}$ transport},}\ }\href
  {\doibase 10.1103/PhysRevLett.95.206603} {\bibfield  {journal} {\bibinfo
  {journal} {Phys. Rev. Lett.}\ }\textbf {\bibinfo {volume} {95}},\ \bibinfo
  {pages} {206603} (\bibinfo {year} {2005})}\BibitemShut {NoStop}%
\bibitem [{\citenamefont {Basko}\ \emph {et~al.}(2006)\citenamefont {Basko},
  \citenamefont {Aleiner},\ and\ \citenamefont {Altshuler}}]{basko2006metal}%
  \BibitemOpen
  \bibfield  {author} {\bibinfo {author} {\bibfnamefont {D.~M.}\ \bibnamefont
  {Basko}}, \bibinfo {author} {\bibfnamefont {I.~L.}\ \bibnamefont {Aleiner}},
  \ and\ \bibinfo {author} {\bibfnamefont {B.~L.}\ \bibnamefont {Altshuler}},\
  }\bibfield  {title} {\enquote {\bibinfo {title} {Metal--insulator transition
  in a weakly interacting many-electron system with localized single-particle
  states},}\ }\href
  {http://www.sciencedirect.com/science/article/pii/S0003491605002630}
  {\bibfield  {journal} {\bibinfo  {journal} {Annals of {P}hysics}\ }\textbf
  {\bibinfo {volume} {321}},\ \bibinfo {pages} {1126} (\bibinfo {year}
  {2006})}\BibitemShut {NoStop}%
\bibitem [{\citenamefont {Oganesyan}\ and\ \citenamefont
  {Huse}(2007)}]{oganesyan2007localisation}%
  \BibitemOpen
  \bibfield  {author} {\bibinfo {author} {\bibfnamefont {V.}~\bibnamefont
  {Oganesyan}}\ and\ \bibinfo {author} {\bibfnamefont {D.~A.}\ \bibnamefont
  {Huse}},\ }\bibfield  {title} {\enquote {\bibinfo {title} {Localization of
  interacting fermions at high temperature},}\ }\href {\doibase
  10.1103/PhysRevB.75.155111} {\bibfield  {journal} {\bibinfo  {journal} {Phys.
  Rev. B}\ }\textbf {\bibinfo {volume} {75}},\ \bibinfo {pages} {155111}
  (\bibinfo {year} {2007})}\BibitemShut {NoStop}%
\bibitem [{\citenamefont {Pal}\ and\ \citenamefont {Huse}(2010)}]{pal2010many}%
  \BibitemOpen
  \bibfield  {author} {\bibinfo {author} {\bibfnamefont {A.}~\bibnamefont
  {Pal}}\ and\ \bibinfo {author} {\bibfnamefont {D.~A.}\ \bibnamefont {Huse}},\
  }\bibfield  {title} {\enquote {\bibinfo {title} {Many-body localization phase
  transition},}\ }\href {\doibase 10.1103/PhysRevB.82.174411} {\bibfield
  {journal} {\bibinfo  {journal} {Phys. Rev. B}\ }\textbf {\bibinfo {volume}
  {82}},\ \bibinfo {pages} {174411} (\bibinfo {year} {2010})}\BibitemShut
  {NoStop}%
\bibitem [{\citenamefont {Nandkishore}\ and\ \citenamefont
  {Huse}(2015)}]{nandkishore2015many}%
  \BibitemOpen
  \bibfield  {author} {\bibinfo {author} {\bibfnamefont {R.}~\bibnamefont
  {Nandkishore}}\ and\ \bibinfo {author} {\bibfnamefont {D.~A.}\ \bibnamefont
  {Huse}},\ }\bibfield  {title} {\enquote {\bibinfo {title} {Many-body
  localization and thermalization in quantum statistical mechanics},}\ }\href
  {\doibase 10.1146/annurev-conmatphys-031214-014726} {\bibfield  {journal}
  {\bibinfo  {journal} {Annu. Rev. Condens. Matter Phys.}\ }\textbf {\bibinfo
  {volume} {6}},\ \bibinfo {pages} {15} (\bibinfo {year} {2015})}\BibitemShut
  {NoStop}%
\bibitem [{\citenamefont {Alet}\ and\ \citenamefont
  {Laflorencie}(2018)}]{alet2018many}%
  \BibitemOpen
  \bibfield  {author} {\bibinfo {author} {\bibfnamefont {F.}~\bibnamefont
  {Alet}}\ and\ \bibinfo {author} {\bibfnamefont {N.}~\bibnamefont
  {Laflorencie}},\ }\bibfield  {title} {\enquote {\bibinfo {title} {Many-body
  localization: an introduction and selected topics},}\ }\href {\doibase
  https://doi.org/10.1016/j.crhy.2018.03.003} {\bibfield  {journal} {\bibinfo
  {journal} {Comptes Rendus Physique}\ }\textbf {\bibinfo {volume} {19}},\
  \bibinfo {pages} {498--525} (\bibinfo {year} {2018})}\BibitemShut {NoStop}%
\bibitem [{\citenamefont {Abanin}\ \emph {et~al.}(2019)\citenamefont {Abanin},
  \citenamefont {Altman}, \citenamefont {Bloch},\ and\ \citenamefont
  {Serbyn}}]{abanin2019colloquium}%
  \BibitemOpen
  \bibfield  {author} {\bibinfo {author} {\bibfnamefont {D.~A.}\ \bibnamefont
  {Abanin}}, \bibinfo {author} {\bibfnamefont {E.}~\bibnamefont {Altman}},
  \bibinfo {author} {\bibfnamefont {I.}~\bibnamefont {Bloch}}, \ and\ \bibinfo
  {author} {\bibfnamefont {M.}~\bibnamefont {Serbyn}},\ }\bibfield  {title}
  {\enquote {\bibinfo {title} {Colloquium: Many-body localization,
  thermalization, and entanglement},}\ }\href {\doibase
  10.1103/RevModPhys.91.021001} {\bibfield  {journal} {\bibinfo  {journal}
  {Rev. Mod. Phys.}\ }\textbf {\bibinfo {volume} {91}},\ \bibinfo {pages}
  {021001} (\bibinfo {year} {2019})}\BibitemShut {NoStop}%
\bibitem [{\citenamefont {Kj\"all}\ \emph {et~al.}(2014)\citenamefont
  {Kj\"all}, \citenamefont {Bardarson},\ and\ \citenamefont
  {Pollmann}}]{kjall2014many}%
  \BibitemOpen
  \bibfield  {author} {\bibinfo {author} {\bibfnamefont {J.~A.}\ \bibnamefont
  {Kj\"all}}, \bibinfo {author} {\bibfnamefont {J.~H.}\ \bibnamefont
  {Bardarson}}, \ and\ \bibinfo {author} {\bibfnamefont {F.}~\bibnamefont
  {Pollmann}},\ }\bibfield  {title} {\enquote {\bibinfo {title} {Many-body
  localization in a disordered quantum ising chain},}\ }\href {\doibase
  10.1103/PhysRevLett.113.107204} {\bibfield  {journal} {\bibinfo  {journal}
  {Phys. Rev. Lett.}\ }\textbf {\bibinfo {volume} {113}},\ \bibinfo {pages}
  {107204} (\bibinfo {year} {2014})}\BibitemShut {NoStop}%
\bibitem [{\citenamefont {Luitz}\ \emph {et~al.}(2015)\citenamefont {Luitz},
  \citenamefont {Laflorencie},\ and\ \citenamefont {Alet}}]{luitz2015many}%
  \BibitemOpen
  \bibfield  {author} {\bibinfo {author} {\bibfnamefont {D.~J.}\ \bibnamefont
  {Luitz}}, \bibinfo {author} {\bibfnamefont {N.}~\bibnamefont {Laflorencie}},
  \ and\ \bibinfo {author} {\bibfnamefont {F.}~\bibnamefont {Alet}},\
  }\bibfield  {title} {\enquote {\bibinfo {title} {Many-body localization edge
  in the random-field {H}eisenberg chain},}\ }\href {\doibase
  10.1103/PhysRevB.91.081103} {\bibfield  {journal} {\bibinfo  {journal} {Phys.
  Rev. B}\ }\textbf {\bibinfo {volume} {91}},\ \bibinfo {pages} {081103}
  (\bibinfo {year} {2015})}\BibitemShut {NoStop}%
\bibitem [{\citenamefont {Vosk}\ \emph {et~al.}(2015)\citenamefont {Vosk},
  \citenamefont {Huse},\ and\ \citenamefont {Altman}}]{vosk2015theory}%
  \BibitemOpen
  \bibfield  {author} {\bibinfo {author} {\bibfnamefont {R.}~\bibnamefont
  {Vosk}}, \bibinfo {author} {\bibfnamefont {D.~A.}\ \bibnamefont {Huse}}, \
  and\ \bibinfo {author} {\bibfnamefont {E.}~\bibnamefont {Altman}},\
  }\bibfield  {title} {\enquote {\bibinfo {title} {Theory of the many-body
  localization transition in one-dimensional systems},}\ }\href {\doibase
  10.1103/PhysRevX.5.031032} {\bibfield  {journal} {\bibinfo  {journal} {Phys.
  Rev. X}\ }\textbf {\bibinfo {volume} {5}},\ \bibinfo {pages} {031032}
  (\bibinfo {year} {2015})}\BibitemShut {NoStop}%
\bibitem [{\citenamefont {Potter}\ \emph {et~al.}(2015)\citenamefont {Potter},
  \citenamefont {Vasseur},\ and\ \citenamefont
  {Parameswaran}}]{potter2015universal}%
  \BibitemOpen
  \bibfield  {author} {\bibinfo {author} {\bibfnamefont {A.~C.}\ \bibnamefont
  {Potter}}, \bibinfo {author} {\bibfnamefont {R.}~\bibnamefont {Vasseur}}, \
  and\ \bibinfo {author} {\bibfnamefont {S.~A.}\ \bibnamefont {Parameswaran}},\
  }\bibfield  {title} {\enquote {\bibinfo {title} {Universal properties of
  many-body delocalization transitions},}\ }\href {\doibase
  10.1103/PhysRevX.5.031033} {\bibfield  {journal} {\bibinfo  {journal} {Phys.
  Rev. X}\ }\textbf {\bibinfo {volume} {5}},\ \bibinfo {pages} {031033}
  (\bibinfo {year} {2015})}\BibitemShut {NoStop}%
\bibitem [{\citenamefont {Goremykina}\ \emph {et~al.}(2019)\citenamefont
  {Goremykina}, \citenamefont {Vasseur},\ and\ \citenamefont
  {Serbyn}}]{goremykina2019analytically}%
  \BibitemOpen
  \bibfield  {author} {\bibinfo {author} {\bibfnamefont {A.}~\bibnamefont
  {Goremykina}}, \bibinfo {author} {\bibfnamefont {R.}~\bibnamefont {Vasseur}},
  \ and\ \bibinfo {author} {\bibfnamefont {M.}~\bibnamefont {Serbyn}},\
  }\bibfield  {title} {\enquote {\bibinfo {title} {Analytically solvable
  renormalization group for the many-body localization transition},}\ }\href
  {\doibase 10.1103/PhysRevLett.122.040601} {\bibfield  {journal} {\bibinfo
  {journal} {Phys. Rev. Lett.}\ }\textbf {\bibinfo {volume} {122}},\ \bibinfo
  {pages} {040601} (\bibinfo {year} {2019})}\BibitemShut {NoStop}%
\bibitem [{\citenamefont {Dumitrescu}\ \emph {et~al.}(2019)\citenamefont
  {Dumitrescu}, \citenamefont {Goremykina}, \citenamefont {Parameswaran},
  \citenamefont {Serbyn},\ and\ \citenamefont
  {Vasseur}}]{dumitrescu2018kosterlitz}%
  \BibitemOpen
  \bibfield  {author} {\bibinfo {author} {\bibfnamefont {P.~T.}\ \bibnamefont
  {Dumitrescu}}, \bibinfo {author} {\bibfnamefont {A.}~\bibnamefont
  {Goremykina}}, \bibinfo {author} {\bibfnamefont {S.~A.}\ \bibnamefont
  {Parameswaran}}, \bibinfo {author} {\bibfnamefont {M.}~\bibnamefont
  {Serbyn}}, \ and\ \bibinfo {author} {\bibfnamefont {R.}~\bibnamefont
  {Vasseur}},\ }\bibfield  {title} {\enquote {\bibinfo {title}
  {Kosterlitz-thouless scaling at many-body localization phase transitions},}\
  }\href {\doibase 10.1103/PhysRevB.99.094205} {\bibfield  {journal} {\bibinfo
  {journal} {Phys. Rev. B}\ }\textbf {\bibinfo {volume} {99}},\ \bibinfo
  {pages} {094205} (\bibinfo {year} {2019})}\BibitemShut {NoStop}%
\bibitem [{\citenamefont {Morningstar}\ and\ \citenamefont
  {Huse}(2019)}]{morningstar2019renormalization}%
  \BibitemOpen
  \bibfield  {author} {\bibinfo {author} {\bibfnamefont {A.}~\bibnamefont
  {Morningstar}}\ and\ \bibinfo {author} {\bibfnamefont {D.~A.}\ \bibnamefont
  {Huse}},\ }\bibfield  {title} {\enquote {\bibinfo {title}
  {Renormalization-group study of the many-body localization transition in one
  dimension},}\ }\href {\doibase 10.1103/PhysRevB.99.224205} {\bibfield
  {journal} {\bibinfo  {journal} {Phys. Rev. B}\ }\textbf {\bibinfo {volume}
  {99}},\ \bibinfo {pages} {224205} (\bibinfo {year} {2019})}\BibitemShut
  {NoStop}%
\bibitem [{\citenamefont {Morningstar}\ \emph {et~al.}(2020)\citenamefont
  {Morningstar}, \citenamefont {Huse},\ and\ \citenamefont
  {Imbrie}}]{morningstar2020manybody}%
  \BibitemOpen
  \bibfield  {author} {\bibinfo {author} {\bibfnamefont {A.}~\bibnamefont
  {Morningstar}}, \bibinfo {author} {\bibfnamefont {D.~A.}\ \bibnamefont
  {Huse}}, \ and\ \bibinfo {author} {\bibfnamefont {J.~Z.}\ \bibnamefont
  {Imbrie}},\ }\bibfield  {title} {\enquote {\bibinfo {title} {Many-body
  localization near the critical point},}\ }\href
  {https://arxiv.org/abs/2006.04825} {\  (\bibinfo {year} {2020})},\ \Eprint
  {http://arxiv.org/abs/2006.04825} {arXiv:2006.04825} \BibitemShut {NoStop}%
\bibitem [{\citenamefont {Logan}\ and\ \citenamefont
  {Wolynes}(1990)}]{logan1990quantum}%
  \BibitemOpen
  \bibfield  {author} {\bibinfo {author} {\bibfnamefont {D.~E.}\ \bibnamefont
  {Logan}}\ and\ \bibinfo {author} {\bibfnamefont {P.~G.}\ \bibnamefont
  {Wolynes}},\ }\bibfield  {title} {\enquote {\bibinfo {title} {Quantum
  localization and energy flow in many-dimensional fermi resonant systems},}\
  }\href {https://aip.scitation.org/doi/10.1063/1.458637} {\bibfield  {journal}
  {\bibinfo  {journal} {J. Chem. Phys.}\ }\textbf {\bibinfo {volume} {93}},\
  \bibinfo {pages} {4994--5012} (\bibinfo {year} {1990})}\BibitemShut {NoStop}%
\bibitem [{\citenamefont {Altshuler}\ \emph {et~al.}(1997)\citenamefont
  {Altshuler}, \citenamefont {Gefen}, \citenamefont {Kamenev},\ and\
  \citenamefont {Levitov}}]{altshuler1997quasiparticle}%
  \BibitemOpen
  \bibfield  {author} {\bibinfo {author} {\bibfnamefont {B.~L.}\ \bibnamefont
  {Altshuler}}, \bibinfo {author} {\bibfnamefont {Y.}~\bibnamefont {Gefen}},
  \bibinfo {author} {\bibfnamefont {A.}~\bibnamefont {Kamenev}}, \ and\
  \bibinfo {author} {\bibfnamefont {L.~S.}\ \bibnamefont {Levitov}},\
  }\bibfield  {title} {\enquote {\bibinfo {title} {Quasiparticle lifetime in a
  finite system: A nonperturbative approach},}\ }\href {\doibase
  10.1103/PhysRevLett.78.2803} {\bibfield  {journal} {\bibinfo  {journal}
  {Phys. Rev. Lett.}\ }\textbf {\bibinfo {volume} {78}},\ \bibinfo {pages}
  {2803--2806} (\bibinfo {year} {1997})}\BibitemShut {NoStop}%
\bibitem [{\citenamefont {Monthus}\ and\ \citenamefont
  {Garel}(2010)}]{MonthusGarel2010PRB}%
  \BibitemOpen
  \bibfield  {author} {\bibinfo {author} {\bibfnamefont {C.}~\bibnamefont
  {Monthus}}\ and\ \bibinfo {author} {\bibfnamefont {T.}~\bibnamefont
  {Garel}},\ }\bibfield  {title} {\enquote {\bibinfo {title} {Many-body
  localization transition in a lattice model of interacting fermions:
  {S}tatistics of renormalized hoppings in configuration space},}\ }\href
  {\doibase 10.1103/PhysRevB.81.134202} {\bibfield  {journal} {\bibinfo
  {journal} {Phys. Rev. B}\ }\textbf {\bibinfo {volume} {81}},\ \bibinfo
  {pages} {134202} (\bibinfo {year} {2010})}\BibitemShut {NoStop}%
\bibitem [{\citenamefont {Pietracaprina}\ \emph {et~al.}(2016)\citenamefont
  {Pietracaprina}, \citenamefont {Ros},\ and\ \citenamefont
  {Scardicchio}}]{pietracaprina2016forward}%
  \BibitemOpen
  \bibfield  {author} {\bibinfo {author} {\bibfnamefont {F.}~\bibnamefont
  {Pietracaprina}}, \bibinfo {author} {\bibfnamefont {V.}~\bibnamefont {Ros}},
  \ and\ \bibinfo {author} {\bibfnamefont {A.}~\bibnamefont {Scardicchio}},\
  }\bibfield  {title} {\enquote {\bibinfo {title} {Forward approximation as a
  mean-field approximation for the {A}nderson and many-body localization
  transitions},}\ }\href {\doibase 10.1103/PhysRevB.93.054201} {\bibfield
  {journal} {\bibinfo  {journal} {Phys. Rev. B}\ }\textbf {\bibinfo {volume}
  {93}},\ \bibinfo {pages} {054201} (\bibinfo {year} {2016})}\BibitemShut
  {NoStop}%
\bibitem [{\citenamefont {Logan}\ and\ \citenamefont
  {Welsh}(2019)}]{logan2019many}%
  \BibitemOpen
  \bibfield  {author} {\bibinfo {author} {\bibfnamefont {D.~E.}\ \bibnamefont
  {Logan}}\ and\ \bibinfo {author} {\bibfnamefont {S.}~\bibnamefont {Welsh}},\
  }\bibfield  {title} {\enquote {\bibinfo {title} {Many-body localization in
  {F}ock space: {A} local perspective},}\ }\href {\doibase
  10.1103/PhysRevB.99.045131} {\bibfield  {journal} {\bibinfo  {journal} {Phys.
  Rev. B}\ }\textbf {\bibinfo {volume} {99}},\ \bibinfo {pages} {045131}
  (\bibinfo {year} {2019})}\BibitemShut {NoStop}%
\bibitem [{\citenamefont {Roy}\ \emph {et~al.}(2019{\natexlab{a}})\citenamefont
  {Roy}, \citenamefont {Logan},\ and\ \citenamefont {Chalker}}]{roy2018exact}%
  \BibitemOpen
  \bibfield  {author} {\bibinfo {author} {\bibfnamefont {S.}~\bibnamefont
  {Roy}}, \bibinfo {author} {\bibfnamefont {D.~E.}\ \bibnamefont {Logan}}, \
  and\ \bibinfo {author} {\bibfnamefont {J.~T.}\ \bibnamefont {Chalker}},\
  }\bibfield  {title} {\enquote {\bibinfo {title} {Exact solution of a
  percolation analog for the many-body localization transition},}\ }\href
  {\doibase 10.1103/PhysRevB.99.220201} {\bibfield  {journal} {\bibinfo
  {journal} {Phys. Rev. B}\ }\textbf {\bibinfo {volume} {99}},\ \bibinfo
  {pages} {220201} (\bibinfo {year} {2019}{\natexlab{a}})}\BibitemShut
  {NoStop}%
\bibitem [{\citenamefont {Roy}\ \emph {et~al.}(2019{\natexlab{b}})\citenamefont
  {Roy}, \citenamefont {Chalker},\ and\ \citenamefont
  {Logan}}]{roy2018percolation}%
  \BibitemOpen
  \bibfield  {author} {\bibinfo {author} {\bibfnamefont {S.}~\bibnamefont
  {Roy}}, \bibinfo {author} {\bibfnamefont {J.~T.}\ \bibnamefont {Chalker}}, \
  and\ \bibinfo {author} {\bibfnamefont {D.~E.}\ \bibnamefont {Logan}},\
  }\bibfield  {title} {\enquote {\bibinfo {title} {Percolation in fock space as
  a proxy for many-body localization},}\ }\href {\doibase
  10.1103/PhysRevB.99.104206} {\bibfield  {journal} {\bibinfo  {journal} {Phys.
  Rev. B}\ }\textbf {\bibinfo {volume} {99}},\ \bibinfo {pages} {104206}
  (\bibinfo {year} {2019}{\natexlab{b}})}\BibitemShut {NoStop}%
\bibitem [{\citenamefont {Roy}\ and\ \citenamefont
  {Logan}(2019)}]{roy2019self}%
  \BibitemOpen
  \bibfield  {author} {\bibinfo {author} {\bibfnamefont {S.}~\bibnamefont
  {Roy}}\ and\ \bibinfo {author} {\bibfnamefont {D.~E.}\ \bibnamefont
  {Logan}},\ }\bibfield  {title} {\enquote {\bibinfo {title} {{Self-consistent
  theory of many-body localisation in a quantum spin chain with long-range
  interactions}},}\ }\href {\doibase 10.21468/SciPostPhys.7.4.042} {\bibfield
  {journal} {\bibinfo  {journal} {SciPost Phys.}\ }\textbf {\bibinfo {volume}
  {7}},\ \bibinfo {pages} {42} (\bibinfo {year} {2019})}\BibitemShut {NoStop}%
\bibitem [{\citenamefont {Pietracaprina}\ and\ \citenamefont
  {Laflorencie}(2019)}]{pietracaprina2019hilbert}%
  \BibitemOpen
  \bibfield  {author} {\bibinfo {author} {\bibfnamefont {F.}~\bibnamefont
  {Pietracaprina}}\ and\ \bibinfo {author} {\bibfnamefont {N.}~\bibnamefont
  {Laflorencie}},\ }\bibfield  {title} {\enquote {\bibinfo {title} {Hilbert
  space fragmentation and many-body localization},}\ }\href@noop {} {\bibfield
  {journal} {\bibinfo  {journal} {arXiv preprint arXiv:1906.05709}\ } (\bibinfo
  {year} {2019})}\BibitemShut {NoStop}%
\bibitem [{\citenamefont {Ghosh}\ \emph {et~al.}(2019)\citenamefont {Ghosh},
  \citenamefont {Acharya}, \citenamefont {Sahu},\ and\ \citenamefont
  {Mukerjee}}]{ghosh2019manybody}%
  \BibitemOpen
  \bibfield  {author} {\bibinfo {author} {\bibfnamefont {S.}~\bibnamefont
  {Ghosh}}, \bibinfo {author} {\bibfnamefont {A.}~\bibnamefont {Acharya}},
  \bibinfo {author} {\bibfnamefont {S.}~\bibnamefont {Sahu}}, \ and\ \bibinfo
  {author} {\bibfnamefont {S.}~\bibnamefont {Mukerjee}},\ }\bibfield  {title}
  {\enquote {\bibinfo {title} {Many-body localization due to correlated
  disorder in fock space},}\ }\href {\doibase 10.1103/PhysRevB.99.165131}
  {\bibfield  {journal} {\bibinfo  {journal} {Phys. Rev. B}\ }\textbf {\bibinfo
  {volume} {99}},\ \bibinfo {pages} {165131} (\bibinfo {year}
  {2019})}\BibitemShut {NoStop}%
\bibitem [{\citenamefont {Roy}\ and\ \citenamefont
  {Logan}(2020)}]{roy2020fock}%
  \BibitemOpen
  \bibfield  {author} {\bibinfo {author} {\bibfnamefont {S.}~\bibnamefont
  {Roy}}\ and\ \bibinfo {author} {\bibfnamefont {D.~E.}\ \bibnamefont
  {Logan}},\ }\bibfield  {title} {\enquote {\bibinfo {title} {Fock-space
  correlations and the origins of many-body localization},}\ }\href {\doibase
  10.1103/PhysRevB.101.134202} {\bibfield  {journal} {\bibinfo  {journal}
  {Phys. Rev. B}\ }\textbf {\bibinfo {volume} {101}},\ \bibinfo {pages}
  {134202} (\bibinfo {year} {2020})}\BibitemShut {NoStop}%
\bibitem [{\citenamefont {Abou-Chacra}\ \emph {et~al.}(1973)\citenamefont
  {Abou-Chacra}, \citenamefont {Thouless},\ and\ \citenamefont
  {Anderson}}]{abou-chacra1973self}%
  \BibitemOpen
  \bibfield  {author} {\bibinfo {author} {\bibfnamefont {R.}~\bibnamefont
  {Abou-Chacra}}, \bibinfo {author} {\bibfnamefont {D.~J.}\ \bibnamefont
  {Thouless}}, \ and\ \bibinfo {author} {\bibfnamefont {P.~W.}\ \bibnamefont
  {Anderson}},\ }\bibfield  {title} {\enquote {\bibinfo {title} {A
  self-consistent theory of localization},}\ }\href {\doibase
  10.1088/0022-3719/6/10/009} {\bibfield  {journal} {\bibinfo  {journal}
  {Journal of Physics C: Solid State Physics}\ }\textbf {\bibinfo {volume}
  {6}},\ \bibinfo {pages} {1734} (\bibinfo {year} {1973})}\BibitemShut
  {NoStop}%
\bibitem [{\citenamefont {Chalker}\ and\ \citenamefont
  {Siak}(1990)}]{chalker1990anderson}%
  \BibitemOpen
  \bibfield  {author} {\bibinfo {author} {\bibfnamefont {J.~T.}\ \bibnamefont
  {Chalker}}\ and\ \bibinfo {author} {\bibfnamefont {S.}~\bibnamefont {Siak}},\
  }\bibfield  {title} {\enquote {\bibinfo {title} {Anderson localisation on a
  {C}ayley tree: a new model with a simple solution},}\ }\href {\doibase
  10.1088/0953-8984/2/11/011} {\bibfield  {journal} {\bibinfo  {journal} {J.
  Phys.: Cond. Matt.}\ }\textbf {\bibinfo {volume} {2}},\ \bibinfo {pages}
  {2671--2686} (\bibinfo {year} {1990})}\BibitemShut {NoStop}%
\bibitem [{\citenamefont {De~Luca}\ \emph {et~al.}(2014)\citenamefont
  {De~Luca}, \citenamefont {Altshuler}, \citenamefont {Kravtsov},\ and\
  \citenamefont {Scardicchio}}]{luca2014anderson}%
  \BibitemOpen
  \bibfield  {author} {\bibinfo {author} {\bibfnamefont {A.}~\bibnamefont
  {De~Luca}}, \bibinfo {author} {\bibfnamefont {B.~L.}\ \bibnamefont
  {Altshuler}}, \bibinfo {author} {\bibfnamefont {V.~E.}\ \bibnamefont
  {Kravtsov}}, \ and\ \bibinfo {author} {\bibfnamefont {A.}~\bibnamefont
  {Scardicchio}},\ }\bibfield  {title} {\enquote {\bibinfo {title} {Anderson
  localization on the {B}ethe lattice: Nonergodicity of extended states},}\
  }\href {\doibase 10.1103/PhysRevLett.113.046806} {\bibfield  {journal}
  {\bibinfo  {journal} {Phys. Rev. Lett.}\ }\textbf {\bibinfo {volume} {113}},\
  \bibinfo {pages} {046806} (\bibinfo {year} {2014})}\BibitemShut {NoStop}%
\bibitem [{\citenamefont {Altshuler}\ \emph {et~al.}(2016)\citenamefont
  {Altshuler}, \citenamefont {Ioffe},\ and\ \citenamefont
  {Kravtsov}}]{altshuler2016multifractal}%
  \BibitemOpen
  \bibfield  {author} {\bibinfo {author} {\bibfnamefont {B.~L.}\ \bibnamefont
  {Altshuler}}, \bibinfo {author} {\bibfnamefont {L.~B.}\ \bibnamefont
  {Ioffe}}, \ and\ \bibinfo {author} {\bibfnamefont {V.~E.}\ \bibnamefont
  {Kravtsov}},\ }\bibfield  {title} {\enquote {\bibinfo {title} {Multifractal
  states in self-consistent theory of localization: analytical solution},}\
  }\href {https://arxiv.org/abs/1610.00758} {\  (\bibinfo {year} {2016})},\
  \Eprint {http://arxiv.org/abs/1610.00758} {arXiv:1610.00758} \BibitemShut
  {NoStop}%
\bibitem [{\citenamefont {Tikhonov}\ \emph {et~al.}(2016)\citenamefont
  {Tikhonov}, \citenamefont {Mirlin},\ and\ \citenamefont
  {Skvortsov}}]{tikhonov2016anderson}%
  \BibitemOpen
  \bibfield  {author} {\bibinfo {author} {\bibfnamefont {K.~S.}\ \bibnamefont
  {Tikhonov}}, \bibinfo {author} {\bibfnamefont {A.~D.}\ \bibnamefont
  {Mirlin}}, \ and\ \bibinfo {author} {\bibfnamefont {M.~A.}\ \bibnamefont
  {Skvortsov}},\ }\bibfield  {title} {\enquote {\bibinfo {title} {Anderson
  localization and ergodicity on random regular graphs},}\ }\href {\doibase
  10.1103/PhysRevB.94.220203} {\bibfield  {journal} {\bibinfo  {journal} {Phys.
  Rev. B}\ }\textbf {\bibinfo {volume} {94}},\ \bibinfo {pages} {220203}
  (\bibinfo {year} {2016})}\BibitemShut {NoStop}%
\bibitem [{\citenamefont {Garc\'{\i}a-Mata}\ \emph {et~al.}(2017)\citenamefont
  {Garc\'{\i}a-Mata}, \citenamefont {Giraud}, \citenamefont {Georgeot},
  \citenamefont {Martin}, \citenamefont {Dubertrand},\ and\ \citenamefont
  {Lemari\'e}}]{garciamata2017scaling}%
  \BibitemOpen
  \bibfield  {author} {\bibinfo {author} {\bibfnamefont {I.}~\bibnamefont
  {Garc\'{\i}a-Mata}}, \bibinfo {author} {\bibfnamefont {O.}~\bibnamefont
  {Giraud}}, \bibinfo {author} {\bibfnamefont {B.}~\bibnamefont {Georgeot}},
  \bibinfo {author} {\bibfnamefont {J.}~\bibnamefont {Martin}}, \bibinfo
  {author} {\bibfnamefont {R.}~\bibnamefont {Dubertrand}}, \ and\ \bibinfo
  {author} {\bibfnamefont {G.}~\bibnamefont {Lemari\'e}},\ }\bibfield  {title}
  {\enquote {\bibinfo {title} {Scaling theory of the {A}nderson transition in
  random graphs: Ergodicity and universality},}\ }\href {\doibase
  10.1103/PhysRevLett.118.166801} {\bibfield  {journal} {\bibinfo  {journal}
  {Phys. Rev. Lett.}\ }\textbf {\bibinfo {volume} {118}},\ \bibinfo {pages}
  {166801} (\bibinfo {year} {2017})}\BibitemShut {NoStop}%
\bibitem [{\citenamefont {Sonner}\ \emph {et~al.}(2017)\citenamefont {Sonner},
  \citenamefont {Tikhonov},\ and\ \citenamefont
  {Mirlin}}]{sonner2017multifractality}%
  \BibitemOpen
  \bibfield  {author} {\bibinfo {author} {\bibfnamefont {M.}~\bibnamefont
  {Sonner}}, \bibinfo {author} {\bibfnamefont {K.~S.}\ \bibnamefont
  {Tikhonov}}, \ and\ \bibinfo {author} {\bibfnamefont {A.~D.}\ \bibnamefont
  {Mirlin}},\ }\bibfield  {title} {\enquote {\bibinfo {title} {Multifractality
  of wave functions on a {C}ayley tree: From root to leaves},}\ }\href
  {\doibase 10.1103/PhysRevB.96.214204} {\bibfield  {journal} {\bibinfo
  {journal} {Phys. Rev. B}\ }\textbf {\bibinfo {volume} {96}},\ \bibinfo
  {pages} {214204} (\bibinfo {year} {2017})}\BibitemShut {NoStop}%
\bibitem [{\citenamefont {Biroli}\ and\ \citenamefont
  {Tarzia}(2018)}]{biroli2018delocalization}%
  \BibitemOpen
  \bibfield  {author} {\bibinfo {author} {\bibfnamefont {G.}~\bibnamefont
  {Biroli}}\ and\ \bibinfo {author} {\bibfnamefont {M.}~\bibnamefont
  {Tarzia}},\ }\bibfield  {title} {\enquote {\bibinfo {title} {Delocalization
  and ergodicity of the {A}nderson model on {B}ethe lattices},}\ }\href
  {https://arxiv.org/abs/1810.07545} {\  (\bibinfo {year} {2018})},\ \Eprint
  {http://arxiv.org/abs/1810.07545} {arXiv:1810.07545} \BibitemShut {NoStop}%
\bibitem [{\citenamefont {Kravtsov}\ \emph {et~al.}(2018)\citenamefont
  {Kravtsov}, \citenamefont {Altshuler},\ and\ \citenamefont
  {Ioffe}}]{kravtsov2018nonergodic}%
  \BibitemOpen
  \bibfield  {author} {\bibinfo {author} {\bibfnamefont {V.~E.}\ \bibnamefont
  {Kravtsov}}, \bibinfo {author} {\bibfnamefont {B.~L.}\ \bibnamefont
  {Altshuler}}, \ and\ \bibinfo {author} {\bibfnamefont {L.~B.}\ \bibnamefont
  {Ioffe}},\ }\bibfield  {title} {\enquote {\bibinfo {title} {Non-ergodic
  delocalized phase in anderson model on bethe lattice and regular graph},}\
  }\href {\doibase https://doi.org/10.1016/j.aop.2017.12.009} {\bibfield
  {journal} {\bibinfo  {journal} {Annals of Physics}\ }\textbf {\bibinfo
  {volume} {389}},\ \bibinfo {pages} {148 -- 191} (\bibinfo {year}
  {2018})}\BibitemShut {NoStop}%
\bibitem [{\citenamefont {Tikhonov}\ and\ \citenamefont
  {Mirlin}(2019)}]{tikhonov2019critical}%
  \BibitemOpen
  \bibfield  {author} {\bibinfo {author} {\bibfnamefont {K.~S.}\ \bibnamefont
  {Tikhonov}}\ and\ \bibinfo {author} {\bibfnamefont {A.~D.}\ \bibnamefont
  {Mirlin}},\ }\bibfield  {title} {\enquote {\bibinfo {title} {Critical
  behavior at the localization transition on random regular graphs},}\ }\href
  {\doibase 10.1103/PhysRevB.99.214202} {\bibfield  {journal} {\bibinfo
  {journal} {Phys. Rev. B}\ }\textbf {\bibinfo {volume} {99}},\ \bibinfo
  {pages} {214202} (\bibinfo {year} {2019})}\BibitemShut {NoStop}%
\bibitem [{\citenamefont {Savitz}\ \emph {et~al.}(2019)\citenamefont {Savitz},
  \citenamefont {Peng},\ and\ \citenamefont {Refael}}]{savitz2019anderson}%
  \BibitemOpen
  \bibfield  {author} {\bibinfo {author} {\bibfnamefont {S.}~\bibnamefont
  {Savitz}}, \bibinfo {author} {\bibfnamefont {C.}~\bibnamefont {Peng}}, \ and\
  \bibinfo {author} {\bibfnamefont {G.}~\bibnamefont {Refael}},\ }\bibfield
  {title} {\enquote {\bibinfo {title} {Anderson localization on the {B}ethe
  lattice using cages and the {W}egner flow},}\ }\href {\doibase
  10.1103/PhysRevB.100.094201} {\bibfield  {journal} {\bibinfo  {journal}
  {Phys. Rev. B}\ }\textbf {\bibinfo {volume} {100}},\ \bibinfo {pages}
  {094201} (\bibinfo {year} {2019})}\BibitemShut {NoStop}%
\bibitem [{\citenamefont {Garc\'{\i}a-Mata}\ \emph {et~al.}(2020)\citenamefont
  {Garc\'{\i}a-Mata}, \citenamefont {Martin}, \citenamefont {Dubertrand},
  \citenamefont {Giraud}, \citenamefont {Georgeot},\ and\ \citenamefont
  {Lemari\'e}}]{garciamata2020two}%
  \BibitemOpen
  \bibfield  {author} {\bibinfo {author} {\bibfnamefont {I.}~\bibnamefont
  {Garc\'{\i}a-Mata}}, \bibinfo {author} {\bibfnamefont {J.}~\bibnamefont
  {Martin}}, \bibinfo {author} {\bibfnamefont {R.}~\bibnamefont {Dubertrand}},
  \bibinfo {author} {\bibfnamefont {O.}~\bibnamefont {Giraud}}, \bibinfo
  {author} {\bibfnamefont {B.}~\bibnamefont {Georgeot}}, \ and\ \bibinfo
  {author} {\bibfnamefont {G.}~\bibnamefont {Lemari\'e}},\ }\bibfield  {title}
  {\enquote {\bibinfo {title} {Two critical localization lengths in the
  {A}nderson transition on random graphs},}\ }\href {\doibase
  10.1103/PhysRevResearch.2.012020} {\bibfield  {journal} {\bibinfo  {journal}
  {Phys. Rev. Research}\ }\textbf {\bibinfo {volume} {2}},\ \bibinfo {pages}
  {012020} (\bibinfo {year} {2020})}\BibitemShut {NoStop}%
\bibitem [{\citenamefont {Tarzia}(2020)}]{tarzia2020manybody}%
  \BibitemOpen
  \bibfield  {author} {\bibinfo {author} {\bibfnamefont {M.}~\bibnamefont
  {Tarzia}},\ }\bibfield  {title} {\enquote {\bibinfo {title} {The many-body
  localization transition in the {H}ilbert space},}\ }\href
  {https://arxiv.org/abs/2003.11847} {\  (\bibinfo {year} {2020})},\ \Eprint
  {http://arxiv.org/abs/2003.11847} {arXiv:2003.11847} \BibitemShut {NoStop}%
\bibitem [{\citenamefont {Biroli}\ and\ \citenamefont
  {Tarzia}(2017)}]{biroli2017delocalized}%
  \BibitemOpen
  \bibfield  {author} {\bibinfo {author} {\bibfnamefont {G.}~\bibnamefont
  {Biroli}}\ and\ \bibinfo {author} {\bibfnamefont {M.}~\bibnamefont
  {Tarzia}},\ }\bibfield  {title} {\enquote {\bibinfo {title} {Delocalized
  glassy dynamics and many-body localization},}\ }\href {\doibase
  10.1103/PhysRevB.96.201114} {\bibfield  {journal} {\bibinfo  {journal} {Phys.
  Rev. B}\ }\textbf {\bibinfo {volume} {96}},\ \bibinfo {pages} {201114}
  (\bibinfo {year} {2017})}\BibitemShut {NoStop}%
\bibitem [{\citenamefont {Biroli}\ and\ \citenamefont
  {Tarzia}(2020)}]{biroli2020anomalous}%
  \BibitemOpen
  \bibfield  {author} {\bibinfo {author} {\bibfnamefont {G.}~\bibnamefont
  {Biroli}}\ and\ \bibinfo {author} {\bibfnamefont {M.}~\bibnamefont
  {Tarzia}},\ }\bibfield  {title} {\enquote {\bibinfo {title} {Anomalous
  dynamics in the ergodic side of the many-body localization transition and the
  glassy phase of directed polymers in random media},}\ }\href
  {https://arxiv.org/abs/2003.09629} {\  (\bibinfo {year} {2020})},\ \Eprint
  {http://arxiv.org/abs/2003.09629} {arXiv:2003.09629} \BibitemShut {NoStop}%
\bibitem [{\citenamefont {De~Tomasi}\ \emph {et~al.}(2020)\citenamefont
  {De~Tomasi}, \citenamefont {Bera}, \citenamefont {Scardicchio},\ and\
  \citenamefont {Khaymovich}}]{detomasi2020subdiffusion}%
  \BibitemOpen
  \bibfield  {author} {\bibinfo {author} {\bibfnamefont {G.}~\bibnamefont
  {De~Tomasi}}, \bibinfo {author} {\bibfnamefont {S.}~\bibnamefont {Bera}},
  \bibinfo {author} {\bibfnamefont {A.}~\bibnamefont {Scardicchio}}, \ and\
  \bibinfo {author} {\bibfnamefont {I.~M.}\ \bibnamefont {Khaymovich}},\
  }\bibfield  {title} {\enquote {\bibinfo {title} {Subdiffusion in the
  {A}nderson model on the random regular graph},}\ }\href {\doibase
  10.1103/PhysRevB.101.100201} {\bibfield  {journal} {\bibinfo  {journal}
  {Phys. Rev. B}\ }\textbf {\bibinfo {volume} {101}},\ \bibinfo {pages}
  {100201} (\bibinfo {year} {2020})}\BibitemShut {NoStop}%
\bibitem [{\citenamefont {Welsh}\ and\ \citenamefont
  {Logan}(2018)}]{welsh2018simple}%
  \BibitemOpen
  \bibfield  {author} {\bibinfo {author} {\bibfnamefont {S.}~\bibnamefont
  {Welsh}}\ and\ \bibinfo {author} {\bibfnamefont {D.~E.}\ \bibnamefont
  {Logan}},\ }\bibfield  {title} {\enquote {\bibinfo {title} {Simple
  probability distributions on a {F}ock-space lattice},}\ }\href
  {http://stacks.iop.org/0953-8984/30/i=40/a=405601} {\bibfield  {journal}
  {\bibinfo  {journal} {J. Phys.: Condens. Matter}\ }\textbf {\bibinfo {volume}
  {30}},\ \bibinfo {pages} {405601} (\bibinfo {year} {2018})}\BibitemShut
  {NoStop}%
\bibitem [{Note1()}]{Note1}%
  \BibitemOpen
  \bibinfo {note} {The algorithm for constructing the correlated energies is
  described in the supplementary material~\cite {supp}}\BibitemShut {NoStop}%
\bibitem [{\citenamefont {Economou}\ and\ \citenamefont
  {Cohen}(1972)}]{economous1972existence}%
  \BibitemOpen
  \bibfield  {author} {\bibinfo {author} {\bibfnamefont {E.~N.}\ \bibnamefont
  {Economou}}\ and\ \bibinfo {author} {\bibfnamefont {M.~H.}\ \bibnamefont
  {Cohen}},\ }\bibfield  {title} {\enquote {\bibinfo {title} {Existence of
  {M}obility {E}dges in {A}nderson's model for {R}andom {L}attices},}\ }\href
  {\doibase 10.1103/PhysRevB.5.2931} {\bibfield  {journal} {\bibinfo  {journal}
  {Phys. Rev. B}\ }\textbf {\bibinfo {volume} {5}},\ \bibinfo {pages}
  {2931--2948} (\bibinfo {year} {1972})}\BibitemShut {NoStop}%
\bibitem [{\citenamefont {Thouless}(1974)}]{ThoulessReview1974}%
  \BibitemOpen
  \bibfield  {author} {\bibinfo {author} {\bibfnamefont {D.~J.}\ \bibnamefont
  {Thouless}},\ }\bibfield  {title} {\enquote {\bibinfo {title} {Electrons in
  disordered systems and the theory of localization},}\ }\href {\doibase
  https://doi.org/10.1016/0370-1573(74)90029-5} {\bibfield  {journal} {\bibinfo
   {journal} {Physics Reports}\ }\textbf {\bibinfo {volume} {13}},\ \bibinfo
  {pages} {93 -- 142} (\bibinfo {year} {1974})}\BibitemShut {NoStop}%
\bibitem [{\citenamefont {Licciardello}\ and\ \citenamefont
  {Economou}(1975)}]{Licciardello+Economou1975}%
  \BibitemOpen
  \bibfield  {author} {\bibinfo {author} {\bibfnamefont {D.~C.}\ \bibnamefont
  {Licciardello}}\ and\ \bibinfo {author} {\bibfnamefont {E.~N.}\ \bibnamefont
  {Economou}},\ }\bibfield  {title} {\enquote {\bibinfo {title} {Study of
  localization in {A}nderson's model for random lattices},}\ }\href {\doibase
  10.1103/PhysRevB.11.3697} {\bibfield  {journal} {\bibinfo  {journal} {Phys.
  Rev. B}\ }\textbf {\bibinfo {volume} {11}},\ \bibinfo {pages} {3697--3717}
  (\bibinfo {year} {1975})}\BibitemShut {NoStop}%
\bibitem [{\citenamefont {Logan}\ and\ \citenamefont
  {Wolynes}(1985)}]{logan1985anderson}%
  \BibitemOpen
  \bibfield  {author} {\bibinfo {author} {\bibfnamefont {D.~E.}\ \bibnamefont
  {Logan}}\ and\ \bibinfo {author} {\bibfnamefont {P.~G.}\ \bibnamefont
  {Wolynes}},\ }\bibfield  {title} {\enquote {\bibinfo {title} {Anderson
  localization in topologically disordered systems},}\ }\href {\doibase
  10.1103/PhysRevB.31.2437} {\bibfield  {journal} {\bibinfo  {journal} {Phys.
  Rev. B}\ }\textbf {\bibinfo {volume} {31}},\ \bibinfo {pages} {2437--2450}
  (\bibinfo {year} {1985})}\BibitemShut {NoStop}%
\bibitem [{\citenamefont {Logan}\ and\ \citenamefont
  {Wolynes}(1987)}]{DELPGWPRB1987}%
  \BibitemOpen
  \bibfield  {author} {\bibinfo {author} {\bibfnamefont {David~E.}\
  \bibnamefont {Logan}}\ and\ \bibinfo {author} {\bibfnamefont {Peter~G.}\
  \bibnamefont {Wolynes}},\ }\bibfield  {title} {\enquote {\bibinfo {title}
  {Dephasing and anderson localization in topologically disordered systems},}\
  }\href {\doibase 10.1103/PhysRevB.36.4135} {\bibfield  {journal} {\bibinfo
  {journal} {Phys. Rev. B}\ }\textbf {\bibinfo {volume} {36}},\ \bibinfo
  {pages} {4135--4147} (\bibinfo {year} {1987})}\BibitemShut {NoStop}%
\bibitem [{Note2()}]{Note2}%
  \BibitemOpen
  \bibinfo {note} {On a tree, there exists a unique shortest path between any
  pair of sites. For a site on generation $l$, the length of the corresponding
  path between it and the root site is $l$.}\BibitemShut {Stop}%
\bibitem [{sup()}]{supp}%
  \BibitemOpen
  \href@noop {} {}\bibinfo {note} {See supplementary material at
  [URL].}\BibitemShut {Stop}%
\bibitem [{Note3()}]{Note3}%
  \BibitemOpen
  \bibinfo {note} {While this analysis focuses on the localised phase,
  self-consistency for the delocalised phase commensurately breaks down at the
  same $W_{c}$ as in Eq.~\protect \textup {\hbox {\mathsurround \z@ \protect
  \normalfont (\ignorespaces \ref {eq:Wc-sc}\unskip \@@italiccorr )}}~\cite
  {logan2019many,roy2019self}.}\BibitemShut {Stop}%
\end{thebibliography}%

\clearpage



\section*{Supplementary material}

## Supplementary material: Localisation on certain graphs with strongly correlated disorder

Sthitadhi Roy and David E. Logan

### Upper-limit approximation

In the recursive formulation for $y_0(\omega)$ discussed in the main text, the effect of the real parts of the self-energies was taken into account exactly, by defining $\Omega_{i_n} = \omega - W\epsilon_{i_n} - X_{i_n}^{(i_{n-1})}(\omega)$ and using the recursion relation for it, Eq. (6). One can however make the 'upper limit approximation' [S1, S2] of neglecting entirely the real parts of the self-energies, setting $X_{i_n}^{(i_{n-1})}(\omega) = 0$. In the case of uncorrelated disorder, this is known to overestimate the critical disorder [S1, S2]. For a tree with maximally correlated disorder Eq. (2), we now show that with this approximation a finite critical disorder is still recovered, albeit substantially larger than that obtained from the recursive formulation. For brevity, we show results for $\omega = 0$ (band centre).

Setting $X_{i_n}^{(i_{n-1})}(\omega) = 0$ leads to a series for $\tilde{y}_0$, similar to Eq. (5), as

$$\tilde{y}_0 = \sum_{i_1 \in \mathsf{N}[0]} \frac{\Gamma^2}{W^2 \epsilon_{i_1}^2} \left[ 1 + \sum_{i_2 \in \mathsf{N}[i_1]} \frac{\Gamma^2}{W^2 \epsilon_{i_2}^2} \left[ 1 + \sum_{i_3 \in \mathsf{N}[i_2]} \frac{\Gamma^2}{W^2 \epsilon_{i_3}^2} [1 + \cdots \right. \right.$$
$$= \sum_l \tilde{\phi}_l; \quad \tilde{\phi}_l = \sum_{i_1 \in \mathsf{N}[0]} \frac{\Gamma^2}{W^2 \epsilon_{i_1}^2} \sum_{i_2 \in \mathsf{N}[i_1]} \frac{\Gamma^2}{W^2 \epsilon_{i_2}^2} \cdots \sum_{i_l \in \mathsf{N}[i_{l-1}]} \frac{\Gamma^2}{W^2 \epsilon_{i_l}^2}, \tag{S1}$$

where we use the notation $\tilde{y}$ and $\tilde{\phi}$ to distinguish upper-limit approximation results from those of the recursive formulation.

For the tree with maximally correlated disorder, $C_{ij} = \exp[-\ell_{ij}/L]$, the distributions of the terms $\tilde{\phi}_l$ and the convergence of the typical value of $\tilde{y}_0^{[l]} \equiv \sum_{n=1}^{l} \tilde{\phi}_n$ obtained from Eq. (S1), can again be studied analogously to that of Fig. 2. Results are shown in Fig. S1, and are qualitatively similar to those of Fig. 2.

At strong disorder the support and the bulk of the distribution of $\tilde{\phi}_l$ move to progressively smaller values with increasing $l$; which manifests itself in the convergence of $\tilde{y}_{0,\mathrm{typ}}^{[l]}$ with $l$, such that $\lim_{L \to \infty} \tilde{y}_{0,\mathrm{typ}}^{[l]}$ is finite. For weak disorder by contrast, the support of the distribution grows with $l$, reflected in an exponential growth of $\tilde{y}_{0,\mathrm{typ}}^{[l]}$ with $l$ such that in the thermodynamic limit $\lim_{L \to \infty} \tilde{y}_{0,\mathrm{typ}}^{[l]}$ diverges, indicating the breakdown of the localised phase.

Finally, one can also calculate $\tilde{y}_{0,\mathrm{typ}} \equiv \tilde{y}_{0,\mathrm{typ}}^{[L]}$ from the series in Eq. (S1), analogously to that shown in Fig. 1(a). Results for different $L$ are shown in Fig. S2. For sufficiently strong disorder, $\tilde{y}_{0,\mathrm{typ}}$ is independent of $L$ indicating a localised phase, whereas for weak disorder it diverges with $L$, symptomatic of a delocalised phase. The critical regime estimated is naturally at larger disorder strength ($W_c \simeq 15.5$) than that obtained from the recursive formalism ($W_c \simeq 6.8$).

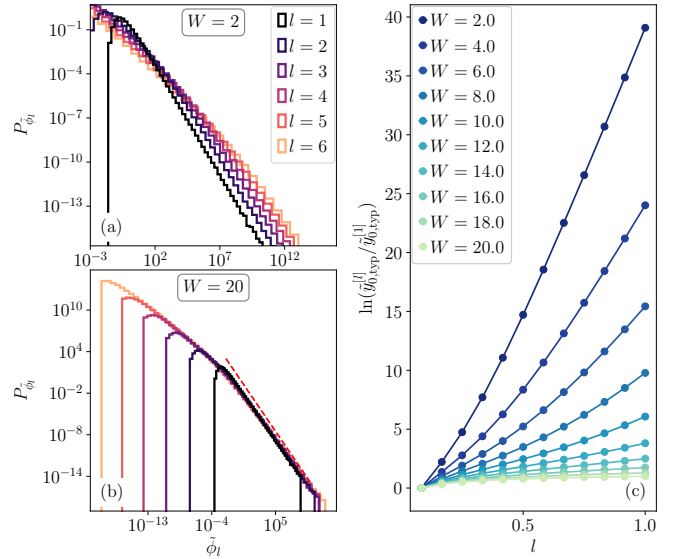

FIG. S1. Convergence of the series for $\tilde{y}_0$, Eq. (S1), under neglect of the real part of the self-energy. Results are qualitatively similar to Fig. 2 (see discussion in main text), except the critical disorder is substantially higher ($W_c \simeq 15.5$). Results shown for $L = 12$ and $10^6$ disorder realisations.

### Wavefunction densities and localisation length

Here we discuss the direct connection between the distribution of the self-energies and that of wavefunction densities in the localised phase; the latter probing the localisation length and its divergence at the transition.

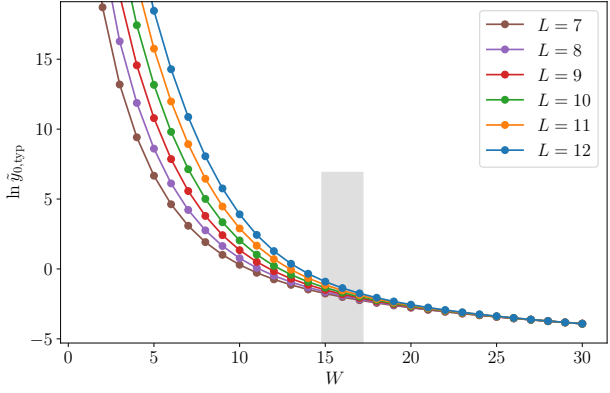

FIG. S2. $\tilde{y}_{0,\mathrm{typ}}$ obtained from the series in Eq. (S1). As in Fig. 1(a), for strong disorder $\tilde{y}_{0,\mathrm{typ}}$ is independent of $L$, while at weak disorder it diverges with $L$. The critical disorder arising from this upper limit approximation is $W_c \simeq 15.5$.

For an eigenstate $|\psi_n\rangle$ of energy $E_n$, the wavefunction density on site $i$ is given (using Eq. (9)) by [S3]

$$w_{ni} \equiv |\langle \psi_n | i \rangle|^2 = [1 + \Delta_i(\omega = E_n)/\eta]^{-1}$$
$$= [1 + y_i(\omega = E_n)]^{-1}. \tag{S2}$$

In the localised phase the distribution of $y$ is a Lévy characterised by a scale parameter $\tilde{\kappa}$,

$$P_y(y) = \sqrt{\tilde{\kappa}/\pi}\; y^{-3/2} e^{-\tilde{\kappa}/y}, \tag{S3}$$

where $\tilde{\kappa} \propto y_{\mathrm{typ}}$ and diverges at the transition.

Transforming the distribution in Eq. (S3) using Eq. (S2), the distribution for $w$ follows,

$$P_w(w) = \sqrt{\frac{\tilde{\kappa}}{\pi}} \frac{1}{w^{1/2}(1-w)^{3/2}} \exp\left[-\frac{\tilde{\kappa}w}{(1-w)}\right]. \tag{S4}$$

From Eq. (S4) the generalised inverse participation ratios (IPRs), $I_q = \sum_i w_{ni}^q$ with $q \geq 1$, can then be obtained as

$$I_q = \frac{\int_0^1 dw\; w^q P_w(w)}{\int_0^1 dw\; w P_w(w)}, \tag{S5}$$

where the denominator takes care of wavefunction normalisation ($I_1 = 1$). Analysis of the integral in Eq. (S5) shows that on approaching the transition from the localised side, $I_q$ vanishes as $\tilde{\kappa}^{1-q}$ for all $q > 1$. In particular, the IPR $I_2$ can be identified as the inverse of the localisation length $\xi$; whence the latter diverges as $\tilde{\kappa}$.

Within the mean-field theory, $\tilde{\kappa} = K\Gamma^2[2(W^2 - 2e^\gamma K\Gamma^2)]^{-1}$ and consequently diverges with an exponent of 1. The mean-field localisation length exponent is thus also unity, $\xi \sim (W - W_c)^{-1}$.

Finally, as noted in the main text, since a Lévy distribution is of one-parameter form (its scale parameter $\tilde{\gamma}$), it can be cast in the scaling form $P_y(y) = y_{\mathrm{typ}}^{-1}\mathcal{P}_{\tilde{y}}(\tilde{y})$;

where the distribution $\mathcal{P}_{\tilde{y}}(\tilde{y})$ of $\tilde{y} = y/y_{\mathrm{typ}}$ is universal ($\mathcal{P}_{\tilde{y}}(\tilde{y}) = [\pi c]^{-1/2}\tilde{y}^{-3/2}\exp(-1/c\tilde{y})$ with $c = 4e^\gamma$ and $\gamma$ Euler's constant). Distributions of $y$ for different $W > W_c$ can thus be collapsed onto a universal form by scaling the self-energy as $y/y_{\mathrm{typ}}$.

**Construction of correlated energies**

We now outline the algorithm for generating the set of correlated site energies. Note that the covariance matrix $\mathbf{C}$ depends solely on the matrix of shortest distances between sites, denoted by $\mathbf{R}$ and obtained using the Floyd-Warshall algorithm for both the Cayley tree and the RRG. From the distance matrix $\mathbf{R}$, the covariance matrix can be trivially obtained element wise as $C_{ij} = f(R_{ij}/L)$. Since $\mathbf{C}$ is a covariance matrix, there exists a similarity transformation which diagonalises it,

$$\mathbf{D} = \mathbf{U}^\mathsf{T}\mathbf{C}\mathbf{U}, \tag{S6}$$

where $\mathbf{D}$ is a diagonal matrix with all non-negative entries. The correlated set of energies, $\{\epsilon_i\}$ can then be generated from an uncorrelated Gaussian random set of energies, $\{e_i\}$ as

$$\epsilon_i = \sum_k [\mathbf{U}\sqrt{\mathbf{D}}]_{ik} e_k. \tag{S7}$$

In our numerical calculations, the system sizes are limited practically by the fact that to generate the correlated energies we need to diagonalise fully the non-sparse matrix $\mathbf{C}$. In principle the recursive formulation can however be implemented for much larger system sizes.

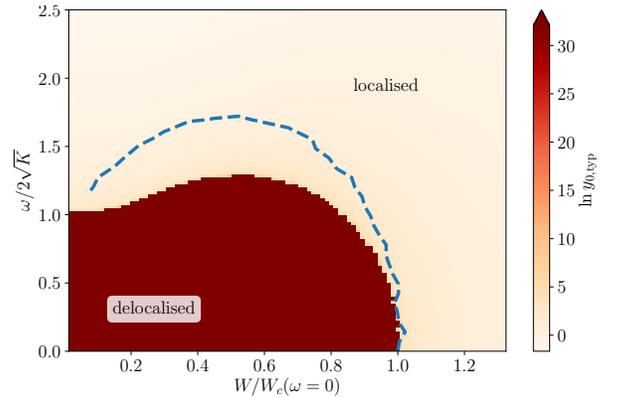

FIG. S3. Mobility edges obtained from the mean-field theory. Colour plot shows $\ln y_{0,\mathrm{typ}}$ obtained self-consistently; blue dashed line shows mobility edges obtained from the recursive formulation (Fig. 1(b)). Note that the disorder strength has been rescaled with the critical value at $\omega = 0$.

## Mean-field mobility edges

The self-consistent mean-field theory was illustrated in the main text for $\omega = 0$. Fig. S3 gives mean-field results arising for $\omega \neq 0$. These too show clear mobility edges, as found from the recursive formulation (Fig. 1(b)), to which it is compared. While the two agree qualitatively, the mean-field theory not unexpectedly overestimates the domain of localised states; as evident both at finite-$\omega$ in general and from the fact that the $\omega = 0$ mean-field $W_c \simeq 2.7$ (Eq. (8)), in contrast to $W_c \simeq 6.8$ from the recursive treatment.

[S1] P. W. Anderson, "Absence of diffusion in certain random lattices," Phys. Rev. **109**, 1492–1505 (1958).

[S2] R. Abou-Chacra, D. J. Thouless, and P. W. Anderson, "A self-consistent theory of localization," Journal of Physics C: Solid State Physics **6**, 1734 (1973).

[S3] D. J. Thouless, "Electrons in disordered systems and the theory of localization," Physics Reports **13**, 93 – 142 (1974).


\end{document}